\documentclass[11pt,a4paper]{article}

\usepackage{fullpage}
\usepackage{amsmath}
\usepackage{amssymb}
\usepackage{amsthm}
\usepackage{mathrsfs}
\usepackage{graphicx}
\usepackage{psfrag}
\usepackage{citesort}
\usepackage{cite} 

\newcommand {\be} {\begin{eqnarray*}}
\newcommand {\ee} {\end{eqnarray*}}
\newcommand {\bea} {\begin{eqnarray}}
\newcommand {\eea} {\end{eqnarray}}

\newcommand{\bm}[1]{\boldsymbol{#1}}

\newcommand{\pdiff}[2]{\frac{\partial{#1}}{\partial{#2}}}

\newcommand{\dblwedge}[0]{\rlap{$\wedge$}\!\wedge}

\newcommand{\dbldif}[0]{\rlap{$\mathrm{d}$}\hspace{0.3ex}\mathrm{d}}

\newcommand{\hateq}[0]{\mathrel{\widehat{=}}}

\newcommand{\qedd}{\hfill $\blacksquare$}
\newcommand{\tr}{{\rm tr}}

\newcommand{\B}[1]{B_{(#1, \ell)}}
\newcommand{\A}[1]{A_{(#1)}}
\newcommand{\areaform}{\tilde{\epsilon}}
\newcommand{\area}{a} 
\newcommand{\IHslice}{\mathcal{S}} 
\newcommand{\slD}{\mathcal{D}}

\newcommand{\pb}[1]{\underleftarrow{#1}}
\newcommand{\undertilde}[1]
  {\mathrm{\raisebox{0.5ex}[0.5ex][-0.5ex]{$\underset{
    \scriptscriptstyle \sim}{}$}}
  \hspace{-0.5em} #1 }
\newcommand{\undersquigarrow}[1]
  {\mathrm{\raisebox{0.65ex}[0.35ex][-0.65ex]{$\underset{
   \!\! {\scriptscriptstyle <} \!\! {\scriptstyle \sim}}{}$}}
  \hspace{-0.5em} #1 }

\newcommand{\gA}{{}^\gamma\!A}
\newcommand{\Hil}{\mathcal{H}}
\newcommand{\phys}{\mathrm{phys}}
\newcommand{\Phys}{\mathrm{Phys}}
\newcommand{\Diff}{\mathrm{Diff}}
\newcommand{\Kin}{\mathrm{Kin}}
\newcommand{\bh}{\mathrm{bh}}
\newcommand{\dirlim}{\underset{\mathcal{P} \subset \IHslice}{\underrightarrow{\lim}}}

\def\R{{\mathbb R}}

\def\D{{D}}

\def\IH{{\Delta}}
\def\={\,\widehat{=}\,}
\def\l{\ell}

\def\d{{\rm d}}

\newcommand{\note}[1]{}

\title{Isolated horizons in classical and quantum gravity\footnote{
Contribution to ``Black Holes: New Horizons'', edited by S. Hayward, to be published by World Scientific}}
\author{\textbf{Jonathan Engle} and \textbf{Tom$\acute{\mbox{a}}\check{\mbox{s}}$ Liko}}

\begin{document}

\maketitle

\section{Introduction}

The classic, global approach to black holes in terms of event horizons has accomplished much: One has a definition of black hole horizon which is
unambiguous and general, depending only on the assumption that space-time is asymptotically flat or asymptotically (anti-)de Sitter.  It is a definition valid for fully dynamical black holes that
possess no symmetry.  For event horizons, one can prove among other things Hawking's very general area theorem, which implies that the areas of event horizons always grow or stay constant in time.  If one furthermore restricts consideration to \textit{stationary} black holes, then the ADM angular momentum and ADM energy can be interpreted as the angular momentum and energy of the black hole itself, giving a notion of  these quantities for black holes that is well-rooted in their deeper meaning as generators of
flows on phase space.
By using the global stationary Killing field available for such space-times, one also arrives at well-defined notions of surface gravity and angular velocity of the horizon. These quantities,
together with surface area, energy and angular momentum, satisfy the zeroth and first laws of black hole mechanics, part of the evidence suggesting that black holes are thermodynamic objects.

In spite of these successes, both of these notions of black hole are, from a physical perspective, not completely satisfactory:
\begin{itemize}
\item The event horizon definition requires knowledge of the entire space-time all the way to future null infinity. However, physically, one can never know the full history of the universe, nor can one measure quantities at infinity.
\item The use of stationary space-times to derive black hole thermodynamics is also not ideal: In all other physical situations, in order to derive laws of equilibrium thermodynamics, it is only necessary to assume equilibrium of the system in question, not the entire universe.
\item Furthermore, beyond such physical considerations, the global nature of event horizons and Killing horizons make them difficult to use in practice.  In particular, in quantum theory, in order for a definition of black hole to make sense, one needs to be able to formulate it in terms of phase space functions which can be quantized.  Event horizons are not amenable to such a characterization in any obvious or simple way. By contrast, the condition defining a Killing horizon can be formulated even in terms of local functions on space (if the Killing field is fixed). However, because the restriction is imposed on fields outside, as well as within, the horizon, the degrees of freedom outside the horizon are reduced and it is again no longer clear if canonical quantum gravity methods, in particular those of loop quantum gravity, can be applied reliably there.
\item Additionally, in numerical relativity, the \textit{global} notions of ADM energy and ADM angular momentum are of limited use, because they do not distinguish the mass of black holes from the energy of surrounding gravitational radiation. If a quasi-local framework could enable a clear definition of angular momentum and energy of a black hole using standard canonical notions, this would be useful for interpreting numerical simulations in a gauge-invariant and systematic way.
\end{itemize}
For all of the above reasons it is both physically desireable
and practical to seek a \textit{quasi-local} alternative to the event horizon and Killing horizon frameworks.

One can ask: Is it possible to find a framework for describing black holes which is quasi-local yet \textit{nevertheless retains all of the desireable features of Killing horizons}?  It is not a priori obvious that this is possible, because part of the reason for the success of the Killing horizon framework is that one can use global stationarity to relate physics at the horizon to physics at infinity, where the asymptotic flat metric is available to aid in the definition of various quantities.  In a quasi-local approach, such access to the structures at infinity will not be possible.  In spite of this challenge, as we shall see, the \textit{isolated horizon} framework allows one to  answer the above question in the affirmative.  Moreover,
due to the fact that the isolated horizon conditions restrict only the \textit{intrinsic geometric structure of the horizon}, they can be implemented in quantum theory,  giving rise to a quantum description of black holes within loop quantum gravity.  In this framework of quantum isolated horizons, one can account for the statistical mechanical origin of the Bekenstein-Hawking entropy of black holes in a large number of physically relevant situations.

The isolated horizon framework has been developed by many authors \cite{abf1998, abf1999, ack1999, ac1999, ashtekar_etal2000, afk2000, abl2001, abl2001a, awd2002, acs2003, aepv2004, lb2007, bl2009, enpp2010}.  Furthermore, the framework is closely related to, and inspired by, not only the Killing horizons as hinted above, but also the trapping horizons of Hayward \cite{hayward2004, hayward1994, hayward1994a}.
Roughly speaking, isolated horizons correspond to portions of trapping horizons that are null. Portions of trapping horizons that are space-like roughly correspond to a complementary notion called `dynamical horizons' \cite{ak2003}.  A more broad review of both of these topics and their relation to trapping horizons has been given in \cite{ak2004}.

In this chapter, after reviewing Killing horizons and black-hole thermodynamics for motivational purposes, we review the geometrical observations regarding null surfaces which naturally lead to the isolated horizon definition.  We then review a covariant canonical framework for space-times with isolated horizons, leading to a canonical, quasi-local notion of angular momentum and mass which then satisfy a \textit{quasi-local} version of the first law of black hole mechanics.   Motivated by the expression for the angular momentum, quasi-local angular momentum and mass multipoles for isolated horizons are also reviewed, in passing.
Finally, we review the quantization of this framework, using the methods of loop quantum gravity
\cite{al2004, rovelli2004, thiemann2007}, leading to a statistical mechanical explanation of the Bekenstein-Hawking entropy of black holes.

\section{Motivation from Killing horizons and black-hole thermodynamics}

\subsection{Black hole horizons and the four laws of black-hole mechanics}
\label{bhmechanics}

A black hole is by definition an object with a gravitational field so strong that no radiation
can escape from it to the asymptotic region of space-time.
Mathematically, a \emph{black hole region} $B$ of a space-time $\mathcal{M}$
is that which excludes all events that belong to the causal past of future null infinity, here denoted $\mathcal{I}^+$:
$B=\mathcal{M}\setminus J^{-}(\mathcal{I}^{+})$, with $J^{-}(N)$ representing the union of the
causal pasts of all events contained in $N$ \cite{he1973}.  The \emph{event horizon}, $\mathcal{H}$, is the
boundary of the black hole region: $\mathcal{H}=\partial B$.  The two-dimensional \emph{cross
section} $\IHslice$ of $\mathcal{H}$ is the intersection of $\mathcal{H}$ with a
three-dimensional spacelike (partial Cauchy) hypersurface $M$: $\IHslice=\mathcal{H}\cap M$.
%
%

A space-time is said to be \emph{stationary} if the metric admits a Killing field $t^{a}$
($a,b,\cdots\in\{0,1,2,3\}$) which approaches a time-translation at spatial infinity.
A space-time is said to be \emph{axi-symmetric} if the metric admits a Killing field $\phi^a$ that generates an $SO(2)$ isometry.
Examples of space-times that are both stationary and axisymmetric are the Schwarzschild solution
and more generally the Kerr-Newman family of solutions. The stationary Killing field $t^a$
is normalized by requiring that it be unit at spatial infinity, while the axisymmetric Killing field $\phi^a$ is normalized by requiring that the affine length of its orbits be $2\pi$.

A space-time is said to contain a \emph{Killing horizon} $\mathcal{K}$ if it possesses
a Killing vector field $\xi^{a}$ that becomes null at $\mathcal{K}$ and generates $\mathcal{K}$.
That is, $\mathcal{K}$ is a null hypersurface with $\xi^a$ as its null normal \cite{wald1994, wald1984}. For stationary space-times, under very general assumptions, the event horizon is
also a Killing horizon \cite[p.331]{he1973}.
%
%
As $\xi^{a}$ is by definition hypersurface-orthogonal at $\mathcal{K}$,
by the Frobenius theorem it satisfies
\begin{displaymath}
\xi_{[a} \nabla_{b} \xi_{c]} \hateq 0
\end{displaymath}
where $\hateq$ denotes equality on $\mathcal{K}$.
From this one can furthermore deduce that
$\xi$ is geodesic at $\mathcal{K}$:
\bea
\xi^{a}\nabla_{a}\xi^{b} \hateq \kappa\xi^{b} \; .
\label{surfaceg}
\eea
This defines the surface gravity $\kappa$ of the black hole.  From this definition it is clear
that the surface gravity rescales as $\kappa\rightarrow c\kappa$ under rescalings of the Killing
vector field $\xi\rightarrow c\xi$, with $c$ a constant.  Equation (\ref{surfaceg}) can
be rewritten
\be
2\kappa\xi_{a} \hateq \nabla_a(-\xi_{b}\xi^{b}) \; .
\ee
Making use of the Frobenius theorem, geodesic equation and Killing equation for the vector
$\xi^{a}$, one obtains the following explicit expression for the surface gravity $\kappa$:
\bea
\kappa^{2} = -\frac{1}{2}(\nabla_a \xi_b)(\nabla^a \xi^b) \; .
\eea
If the dominant energy condition is satisfied, then it follows that $\kappa$ not only is
constant along the null generators of $\mathcal{H}$, but also it does not vary from generator
to generator.  This means that $\kappa$ remains constant over all of $\mathcal{H}$.  This
is the zeroth law of black-hole mechanics:

\noindent{\bf Zeroth Law.}
\emph{If the dominant energy condition is satisfied, then the surface gravity $\kappa$ is
constant over the entire event horizon $\mathcal{H}$.}

The black-hole uniqueness theorems
\cite{israel1968, carter1971, he1973, carter1973, robinson1975, mazur1982}
state that there is a unique three-parameter set of
solutions to the Einstein-Maxwell field equations that are stationary and asymptotically flat;
these are the Kerr-Newman family of solutions parameterized by the ADM mass $M$, ADM
angular momentum $J$ and electric charge $Q$ of the space-time.
For this family, the Killing vector field
defining $\mathcal{K}$ is given by $\xi^{a}=t^{a}+\Omega\phi^{a}$, where $\Omega$ is called
the angular velocity of $\mathcal{K}$.  The electric potential at $\mathcal{K}$ is furthermore
defined as $\Phi=-\bm{A}_{a}\xi^{a}|_{\mathcal{K}}$ where and $\bm{A}_{a}$
is the 4-vector potential of the electromagnetic field.
If the mass of a solution contained in the Kerr-Newman family
is perturbed by an amount $\delta M$, then the changes in the surface area $\area$ and asymptotic
charges $(M,Q,J)$ of $\mathcal{K}$ are governed by the following law.

\noindent{\bf First Law.}
\emph{Let $\area$, $\kappa$, $\Omega$, $\Phi$, and $(M,J,Q)$ denote the horizon area, surface gravity, angular velocity,
electric potential, and asymptotic charges of a stationary black hole.
For any variation $\delta$ within the space of stationary
black holes, one has}
\bea
\delta M = \frac{\kappa}{8\pi G}\delta \area + \Phi\delta Q + \Omega\delta J \; .
\label{firstlaw2}
\eea

This form of the first law is known as the ``equilibrium'' form.  That is, the condition
(\ref{firstlaw2}) describes the changes of the black-hole parameters from a solution to
\emph{nearby solutions} within the phase space.  There is, however, a ``physical process''
interpretation: if one drops a small mass $\delta M$ of matter into a black hole, the resulting
changes $\delta Q$ and $\delta J$ in the charge and angular momentum of the black hole will be
such that the first law equation is satisfied. Such an interpretation is valid if the space-time
is quasi-stationary --- i.e., approximately stationary at each point in time.

If some matter with stress energy $T_{ab}$ is dropped into a black hole, then the mass of
black hole is going to change.  From the first law, the surface area $\area$ will have to
change as well.  A remarkable fact is that, if $T_{ab}\xi^{a}\xi^{b}\ge 0$, then
this change cannot be a decrease.  This is a statement of the second law of black-hole
mechanics:

\noindent{\bf Second Law}
\emph{The surface area $\area$ can never decrease in a physical process if the
stress-energy tensor $T_{ab}$ satisfies the null energy condition $T_{ab}\xi^{a}\xi^{b}\geq0$.}

One final property of quasi-stationary horizons is that it is impossible for one
to become extremal within finite advanced time (i.e., finite time as experienced by free-falling
observers near the horizon) \cite{israel1986}.
%
%
That is,
the surface gravity cannot be reduced to zero in finite advanced time.
This is the statement of the third law of black-hole mechanics:

\noindent{\bf Third Law.}
\emph{The surface gravity $\kappa$ of a quasi-stationary black hole cannot be reduced to zero by any
physical process within finite advanced time.}

\subsection{Black-hole thermodynamics}

At this point it is instructive to pause for a moment to summarize the established
four laws of black-hole mechanics, and compare them to the corresponding four
laws of thermodynamics.  In the following Table \ref{similarities}, $T$ is the temperature
of a system, $U$ its internal energy and $S$ its entropy.
\begin{table}[t]
\begin{tabular*}{\linewidth}{@{\extracolsep{\fill}}ccc}
\hline
\hline
\quad Law & \quad Thermodynamics & \quad Black holes \\
\hline
& & \\
\quad Zeroth & \quad $T$ constant throughout body & \quad $\kappa$ constant over horizon of \\
\quad \phantom{a} & \quad in thermal equilibrium & \quad a stationary black hole \\
& & \\
\quad First & \quad $dU=TdS+\mbox{work terms}$ &
       \quad $dM=\frac{\kappa}{8\pi}d\area+\Omega_{H}dJ+\Phi_{H}dQ$ \\
& & \\
\quad Second & \quad $\Delta S\geq0$ in any process &
         \quad $\Delta \area\geq0$ in any process \\
& & \\
\quad Third & \quad Impossible to achieve $T=0$ & \quad Impossible to achieve $\kappa=0$ \\
\quad \phantom{a} & \quad by a physical process & \quad by a physical process \\
& & \\
\hline
\hline
\end{tabular*}
\caption{A summary of the four laws of black-hole mechanics and the corresponding laws of
thermodynamics.  Here, we make the identifications $U=M$, $T=\kappa/(2\pi)$ and $S=\area/4$.
[Adapted from \cite{wald1984}.]}
\label{similarities}
\end{table}
As one can see, there is a striking formal similarity between these two sets of laws.  Motivated by this
similarity, Bekenstein conjectured that for a black hole \cite{bekenstein1973,bekenstein1974}:
\be
\kappa \propto T
\quad
\mbox{and}
\quad
\area \propto S \; .
\ee
This identification of the surface area of the horizon with thermal entropy also offered a
way to compensate for the apparent violation of the second law of thermodynamics which would seem to
occur when matter falls into a black hole. The idea is that, because black holes are now endowed with
entropy, a \emph{generalized second law} of thermodynamics would still hold, even during such processes:
\be
\delta S_{\rm Universe} + \delta S_{\rm Black-hole} \geq 0 \; .
\ee
However, it turns out that when trying to match $\area$ with $S$, and in order for $S$ to remain dimensionless, we require a combination of the physical constants $c$, $G$ \emph{and} $\hbar$.  By convention we take
temperature to be measured in units of energy so that the Boltzmann constant is unity, whence the unique
combinations of fundamental constants which will fit into the proportionality relations are
\be
S = \mbox{constant} \times \frac{\area}{4l_{\rm P}^{2}}
\quad
\mbox{and}
\quad
T = \mbox{constant} \times \hbar\kappa \, ,
\ee
with $l_{\rm P}=\sqrt{G\hbar/c^{3}}$ the Planck length.

In 1974, Hawking \cite{hawking1972} discovered that black holes radiate a blackbody spectrum with
a temperature of $T=\hbar\kappa/(2\pi)$.  This is known as the Hawking Effect.  The result came
from considering quantum field theory on a fixed black hole background space-time. Hawking was
able to use this result to fix the proportionality constant in Bekenstein's
surface-gravity/temperature relation by requiring that
\bea
T \delta S = \frac{\kappa}{8\pi G}\delta \area \; .
\eea
One finds that
\bea
S \equiv \frac{\area}{4l_{\rm P}^{2}}
\label{entropy1}
\eea
in Planck units with $\l_{\rm P}=\sqrt{G\hbar/c^{3}}$ the Planck length, or
\bea
S \equiv \frac{\area k_{\rm B}c^{3}}{4\hbar G}
\label{entropy2}
\eea
in SI units with $k_{\rm}$ Boltzmann's constant.

The four laws were first formulated for stationary space-times in four-dimensional
Einstein-Maxwell theory \cite{bch1973}, but later were extended using covariant phase space
methods to include stationary black holes in arbitrary diffeomorphism-invariant theories
\cite{wald1993,iw1994,jkm1994,jkm1995}.  This work revealed, remarkably, that the zeroth law
holds for \emph{any} stationary black hole space-time if the matter fields satisfy an appropriate
energy condition.  In addition, the surface-area term in the first law is modified only in cases
when gravity is supplemented with nonminimally coupled matter or higher-curvature interactions.
For a general Lagrangian density $L=L(g_{ab},\nabla_{c}g_{ab}, R_{abcd}, \nabla_{e}R_{abcd})$
(where $L(\cdot, \cdot, \cdot, \cdot)$ involves no derivatives of its arguments),
the entropy of a stationary black hole space-time is given by
\bea
S = -2\pi\oint_{\IHslice}\frac{\delta L}{\delta R_{abcd}}n_{ab}n_{cd} \, ,
\label{entropy3}
\eea
with $n_{ab}$ the binormal to the cross-section $\IHslice$ of the horizon, with
normalization $n_{ab}n^{ab}=-2$.
The fact that higher-order terms in the curvature affect the entropy of black-hole space-times is
a consequence of the fact that such terms modify the gravitational surface term in the symplectic
structure.  As an example, consider a modification to the Einstein-Hilbert Lagrangian
consisting in adding the Euler density
\bea
\mathcal{L}_{\chi} = R^{2} - 4R_{ab}R^{ab} + R_{abcd}R^{abcd} \, .
\eea
The resulting extra
term in the action is a topological invariant of $\mathcal{M}$, and thus only serves to shift the
value of the action by number which is locally constant in the space of histories.  Nevertheless, it
contributes to the gravitational surface term in the symplectic structure and therefore also shifts the value of the entropy by a
number depending on the topology of $\mathcal{M}$.  Though usually non-dynamical, for black-hole mergers,
this term will in general be dynamical, due to the fact that the topology of $\mathcal{M}$ changes during the merging process
\cite{jm1993,liko2007}.

\subsection{Global equilibrium: limitations}

The standard definition of a black hole event horizon
is teleological in the sense that we need to know the structure
of the entire space-time in order to construct the event horizon.  This is a major drawback.
The usual way to resolve this
is to consider solutions to the field equations that are stationary, as we have done above. These are solutions that
admit a time translation Killing field \emph{everywhere}, not just in a small neighborhood of
the black hole region.  While this simple idealization is a natural starting point, it seems
to be overly restrictive.

\begin{figure}[t]
\begin{center}
\psfrag{H}{$\Delta$}
\psfrag{Ip}{$\mathcal{I}^{+}$}
\psfrag{i0}{$i^{0}$}
\psfrag{S1}{$M_{1}$}
\psfrag{S2}{$M_{2}$}
\psfrag{S}{$\mathscr{S}$}
\psfrag{R}{$\mathscr{R}$}
\includegraphics[width=2in]{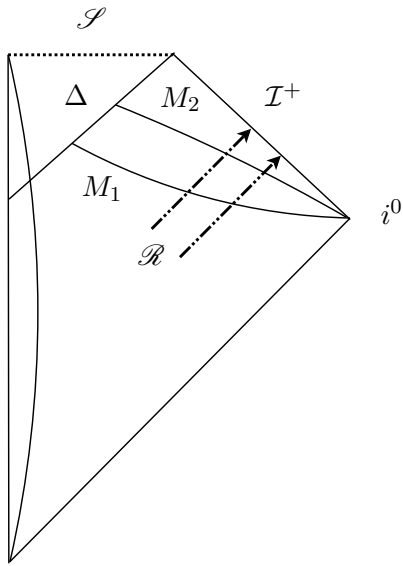}
\caption{A Penrose-Carter diagram of gravitational collapse of an object that forms a
singularity $\mathscr{S}$ and a horizon $\Delta$, in the presence of external matter
fields.  The region of the space-time $\mathcal{M}$ being considered contains the
quasi-local black-hole boundary $\Delta$ intersecting the partial Cauchy surfaces
$M_{1}$ and $M_{2}$ that extend to spatial infinity $i^{0}$.  Here, $\Delta$
is in equilibrium with dynamical radiation fields $\mathscr{R}$ in the exterior region
$\mathcal{M}\setminus\Delta$; these fields are not just in the exterior but can exist
within an arbitrarily small neighbourhood of $\Delta$. \label{pcdiagram}}
\end{center}
\end{figure}

Physically, it should be sufficient to impose boundary conditions at the horizon which ensure
that \emph{only the black hole itself is in equilibrium}.
This viewpoint is consistent with what is usually done in thermodynamics: for the laws of equilibrium thermodynamics to hold in other situations, one need only assume the system in question is in equilibrium, not the whole universe.
 An approach to quasi-local black-hole horizons that achieves this
is the \textit{isolated horizon} (IH) framework.
More precisely, an isolated horizon models a portion of an event horizon
in which the intrinsic geometric structures are `time independent',
and in this sense are in `equilibrium', while the geometry outside may be dynamical, even in
an arbitrarily small neighborhood of the horizon.
In terms of physical processes, an isolated horizon can be characterized as having \textit{no flux of matter or gravitational energy} through it.
In realistic situations of gravitational collapse, such an assumption will be approximately valid only for certain finite
intervals of time, as represented in figure \ref{pcdiagram}.
In the following sections, we define and review the isolated horizon framework in detail, and review
a selection of its applications.

\section{Isolated horizons}

\subsection{Null hypersurfaces in equilibrium: non-expanding horizons}

A null surface $\mathcal{N}$ has a normal $\ell_a$, which, when raised with the space-time metric, is
\textit{tangent} to $\mathcal{N}$.  Given $\mathcal{N}$, the null normal is of course not uniquely
determined, but rather one has the freedom to rescale by a positive smooth function: $\ell_a \mapsto \ell_a^\prime = f \ell_a$.
The intrinsic metric $q_{ab}$ on $\mathcal{N}$ is the pullback of the space-time metric;
because $q_{ab}\ell^{a}\hateq g_{\underleftarrow{ab}}\ell^{a}\hateq\ell_{\underleftarrow{b}}\hateq0$ (with ``$\hateq$''
denoting equality restricted to $\mathcal{N}$ and ``$\underleftarrow{\cdot}$'' denoting pullback to $\mathcal{N}$)
it follows that $q_{ab}$ is degenerate, i.e. $l^{a}$ is the degenerate direction of $q_{ab}$.  The signature of
$q_{ab}$ is then $(0,+,+)$ which means that the determinant of $q_{ab}$ is zero.  As a result, $q_{ab}$ is
non-invertable.  However, an inverse metric $q^{ab}$ can be defined such that $q^{ab}q_{ac}q_{bd} \hateq q_{cd}$;
\emph{any tensor} that satisfies this identity is said to be an inverse of $q_{ab}$.

If $q^{ab}$ is an inverse, then so is $\tilde{q}^{ab}=q^{ab}+\ell^{(a}X^{b)}$ for any $X^b$ tangent to $\mathcal{N}$.
To see this, we observe that
\be
\tilde{q}^{ab} q_{am} q_{bn} &\hateq& (q^{ab} + \ell^{(a}X^{b)})q_{am} q_{bn}\\
                             &\hateq& q^{ab}q_{am}q_{bn} + \frac{1}{2}\ell^{a}X^{b}q_{am}q_{bn}
                                      + \frac{1}{2}\ell^{b}X^{a}q_{am}q_{bn}\\
                             &\hateq& q_{mn} \; .
\ee
In going from the second to the third line, we used the property that $\ell^{a}q_{ab}=0$.

Because $\ell_a$ is hypersurface orthogonal, it satisfies $\pb{\ell_{[a} \nabla_b \ell_{c]}} = 0$.  From this
alone, one can furthermore deduces that $\ell^a$ is geodesic, so that
$\ell^a$ generates a \textit{geodesic null congruence} on $\mathcal{N}$.
The twist, expansion, and shear of this null congruence are given respectively by the
anti-symmetric, trace, and trace-free symmetric parts of $\pb{\nabla_a \ell_b}$. Because $\ell_a$ is
surface-forming, it is twist free.  This leaves the expansion and shear,
\be
\theta_{(\ell)} = q^{ab}\nabla_{a} l_{b}
\quad
\mbox{and}
\quad
\sigma_{ab} = \underleftarrow{\nabla_{(a}\ell_{b)}} - \frac{1}{2}q_{ab}q^{cd}\nabla_{c} l_{d} \, ;
\ee
These both vanish iff $\underleftarrow{\nabla_{a}\ell_{b}}$ vanishes.

An important property of the expansion and shear is that both are independent of the choice
of inverse metric $q^{ab}$.  This follows from the following:
\be
(\tilde{q}^{ab} - q^{ab})\nabla_{a}\ell_{b}
&\hateq& \ell^{(a}X^{b)}\nabla_{a}\ell_{b}\\
&\hateq& \frac{1}{2}[\ell^{a}(\nabla_{a}\ell_{b})X^{b} + (\ell^{b}\nabla_{a}\ell_{b})X^{a}]\\
&\hateq& \frac{1}{2}\kappa_{(\ell)}\ell_{a}X^{a} + \frac{1}{4}X^{a}\nabla_{a}(\ell_{b}\ell^{b}) = 0 \, ;
\ee
the first term vanishes because $\ell_{a}X^{a}=0$ and the second term vanishes because
$\ell_{a}\ell^{a}=0$.

A Killing horizon for stationary space-times is a null hypersurface, and so its null normal generates
a twist-free geodesic null congruence on the horizon.  This null congruence is furthermore \textit{expansion-free}.
The condition that a general null hypersurface $\mathcal{N}$
be expansion free is in fact independent of the null normal used to define the expansion:
If $\ell' = f \ell$ are two null normals related by some positive, smooth function $f$, we have
\bea
\theta_{(\ell^{\prime})} \hateq q^{ab}\nabla_{a}f\ell_{b} \hateq q^{ab}\ell_{b}\nabla_{a}f
                                 + f\theta_{(\ell)} \hateq f\theta_{(\ell)} \; .
\eea
So that $\theta_{(\ell^\prime)} = 0$ iff $\theta_{(\ell)} = 0$. The
vanishing of the expansion is therefore \textit{an intrinsic property of a null surface} $\mathcal{N}$.
%
%
%
This enables us to incorporate this \emph{local} property of a Killing horizon into the following definition.

\noindent{\bf Definition 1. (Non-Expanding Horizon).}
\emph{A three-dimensional null hypersurface $\Delta\subset\mathcal{M}$ of a space-time $(\mathcal{M},g_{ab})$
is said to be a non-expanding horizon (NEH) if the following conditions hold: (i) $\Delta$ is topologically
$R\times \IHslice$ with $\IHslice$ a compact two-dimensional manifold; (ii) the expansion $\theta_{(\ell)}$
of any null normal $\ell$ to $\Delta$ vanishes; (iii) the field equations hold at $\Delta$;
and (iv) the stress-energy tensor $T_{ab}$ of external matter fields is such that, at $\Delta$, $-T^{a}_{b}\ell^{b}$
is a future-directed and causal vector for any future-directed null normal $\ell$.}

For now we leave the horizon cross section $\IHslice$ arbitrary.  In Section \ref{topsect} we will prove that, once the definition
is strengthened a bit more, for
vanishing cosmological constant, $\IHslice$ has to be a two-sphere, thereby generalizing the Hawking
Topology Theorem to non-stationary space-times.

The weakest notion of equilibrium for a NEH is the requirement that $\pounds_\ell q_{ab}\hateq0$.  This means that
the intrinsic geometry of $\Delta$ is invariant under time translations.  The condition is equivalent to
\bea
\pounds_\ell q_{ab} \hateq \underleftarrow{\pounds_\ell g_{ab}}=2\underleftarrow{\nabla_{(a}\ell_{b)}} \hateq 0 \, . \label{qsymm}
\eea
If $\pounds_\ell q_{ab}\hateq0$ for one null normal $\ell$ then it is true for any other null normal $\ell' = f \ell$:
\be
\pounds_{\ell^{\prime}}q_{ab} &\hateq& 2 \pb{\nabla_{(a} (\ell^\prime_{b)})}
\hateq 2 \pb{\nabla_{(a} (f \ell_{b)})} \hateq 2 \pb{(\nabla_{(a} f) \ell_{b)}} + 2 f \pb{\nabla_{(a} \ell_{b)}}
\hateq 2 f \pb{\nabla_{(a} \ell_{b)}} \hateq f \pounds_\ell q_{ab} \; .
\ee
It follows that $\pounds_{\ell^{\prime}}q_{ab}=0$ if $\pounds_{\ell}q_{ab}=0$.

\subsection{Intrinsic geometry of non-expanding horizons}

Let us now discuss the restrictions on the Riemann curvature tensor for space-times in the presence of
a NEH $\Delta$.  The Riemann tensor is defined by the condition $2\nabla_{[a}\nabla_{b]}X^{c}=-R_{abd}{}^{c}X^{d}$;
the tensor $R_{abcd}$ decomposes into a trace part determined by the Ricci tensor $R_{ab}=R_{acb}{}^{c}$ and a
trace-free part $C_{abcd}$ such that:
\bea
R_{abcd} = C_{abcd}
   + \frac{2}{D-2}\left(g_{a[c}R_{d]b} - g_{b[c}R_{d]a}\right)
   - \frac{2}{(D-1)(D-2)}Rg_{a[c}g_{d]b} \; .
\label{decomposition}
\eea
The tensor $C_{abcd}$ is called the Weyl tensor.  The Ricci tensor is determined by the matter fields through the
Einstein-Maxwell equations.  The remaining trace-free part of $R_{abcd}$ therefore corresponds to the gravitational
degrees of freedom.

Next, let us introduce a null basis adapted to $\Delta$.  This can be done by partially gauge-fixing the tetrad so that
$(e^a_0 + e^a_1)/\sqrt{2}$ is a null normal to $\Delta$; we then define
\begin{eqnarray}
\nonumber
\ell^a := \frac{1}{\sqrt{2}}(e^a_0 + e^a_1) &\quad& n^a := \frac{1}{\sqrt{2}}(e^a_0 - e^a_1) \\
m^a := \frac{1}{\sqrt{2}}(e^a_2 + i e^a_3) &\quad& \overline{m}^a := \frac{1}{\sqrt{2}}(e^a_2 - i e^a_3) \; .
\label{npdef}
\end{eqnarray}
These are all null, and satisfy the usual normalizations $\ell^a n_a = 1, m^a \overline{m}_a = 1$ for
a complex Newman-Penrose tetrad.
In terms of this basis, the Riemann tensor can be decomposed into 15 scalar quantities:
\be
\begin{array}{rcl@{\hspace{5mm}}rcl@{\hspace{5mm}}rcl}
    \Psi_{0}  &=& C_{abcd}\ell^{a}m^{b}\ell^{c}m^{d} &
    \Phi_{00} &=& \frac{1}{2}R_{ab}\ell^{a}\ell^{b} &
    \Phi_{12} &=& \frac{1}{2}R_{ab}m^{a}n^{b} \\[1.5mm]
    \Psi_{1}  &=& C_{abcd}\ell^{a}m^{b}\ell^{c}n^{d} &
    \Phi_{01} &=& \frac{1}{2}R_{ab}\ell^{a}m^{b} &
    \Phi_{20} &=& \frac{1}{2}R_{ab}\bar{m}^{a}\bar{m}^{b}  \\[1.5mm]
    \Psi_{2}  &=& C_{abcd}\ell^{a}m^{b}\bar{m}^{c}n^{d} &
    \Phi_{02} &=& \frac{1}{2}R_{ab}m^{a}m^{b} &
    \Phi_{21} &=& \frac{1}{2}R_{ab}\bar{m}^{a}n^{b} \\[1.5mm]
    \Psi_{3}  &=& C_{abcd}\ell^{a}n^{b}\bar{m}^{c}n^{d} &
    \Phi_{10} &=& \frac{1}{2}R_{ab}\ell^{a}\bar{m}^{b} &
    \Phi_{22} &=& \frac{1}{2}R_{ab}n^{a}n^{b} \\[1.5mm]
    \Psi_{4}  &=& C_{abcd}\bar{m}^{a}n^{b}\bar{m}^{c}n^{d} &
    \Phi_{11} &=& \frac{1}{4}R_{ab}(\ell^{a}n^{b} + m^{a}\bar{m}^{b}) &
    \Lambda   &=& \frac{R}{24} \, .
\end{array}
\ee
The four real scalars ($\Phi_{00},\Phi_{11},\Phi_{22},\lambda$) and three complex scalars ($\Phi_{10},\Phi_{20},\Phi_{21}$)
correspond to the Ricci tensor, and the five complex scalars ($\Psi_{0},\Psi_{1},\Psi_{2},\Psi_{3},\Psi_{4}$) correspond
to the Weyl tensor.  Of these, $\Phi_{00}$, $\Phi_{01}$, $\Phi_{10}$, $\Phi_{02}$, $\Phi_{20}$, $\Psi_{0}$ and $\Psi_{1}$ are all
identically zero at $\Delta$ by the boundary conditions and the Raychaudhuri equation;
in particular, the vanishing of $\Psi_0$ and $\Psi_1$ imply that the Weyl tensor is algebraically special at the horizon.
The remaining components of $R_{abcd}$ are
unconstrained at $\Delta$.  Furthermore, one can show that $\Psi_2$ is independent of the null tetrad provided that $\ell$ is a null normal
to $\Delta$.

The above properties of a NEH furthermore ensure that the space-time derivative operator $\nabla_a$
induces a natural \textit{intrinsic} derivative operator on $\Delta$. On a space-like or time-like
hypersurface, an intrinsic derivative operator is \textit{always} induced by $\nabla_a$, but on a null hypersurface
this is not always the case. Recall that a covariant derivative operator $\mathcal{D}$ intrinsic to a manifold $\Sigma$
can be specified by giving a map from each pair of vector fields $X,Y$ on $\Sigma$ to a vector field
$\mathcal{D}_X Y \equiv X^b \mathcal{D}_b Y^a$, also on $\Sigma$, satisfying the axioms
\begin{enumerate}
\item
$\mathcal{D}_Z (X+Y)=D_Z X +D_Z Y$ \, ,
\item
$\mathcal{D}_Z (fX)=fD_Z X +X \pounds_Z f$ \, .
\end{enumerate}
The action of $\mathcal{D}$ on tensors of other types is then determined by linearity $\mathcal{D}_X (T^\mathcal{A} + S^{\mathcal{A}}) =
\mathcal{D}_X T^{\mathcal{A}} + \mathcal{D}_X S^{\mathcal{A}}$
and the Leibnitz rule $\mathcal{D}_X(T^\mathcal{AC}S_\mathcal{BC}) = (\mathcal{D}_X T^\mathcal{AC}) S_\mathcal{BC}
+ T^\mathcal{AC} (\mathcal{D}_X S_\mathcal{BC})$, where $\mathcal{A}, \mathcal{B}, \mathcal{C}$ each denote
any combination of tensor indices.
%
%
%

Given a space-like or time-like hypersurface $\Sigma$, the derivative operator $\mathcal{D}$ induced by
$\nabla_a$ is usually defined by
\begin{displaymath}
(\mathcal{D}_X Y)^a = h^a_b (\nabla_X Y)^b
\end{displaymath}
where $h_a^b := \delta^b_a + s n^b n_a$ is the projector onto the tangent space of $\Sigma$, with $n^a$ a unit normal to $\Sigma$
satisfying $n^a n_a \equiv s = \pm 1$.  On a \textit{null} hypersurface such as $\IH$, however,
no such canonical projector exists, and so the above standard prescription fails.
In order for $\nabla_a$ to induce a natural derivative operator on $\Delta$, one therefore needs the property that,
if $X$ and $Y$ are vector fields tangent to $\Delta$, then $\nabla_X Y$ is \textit{already also} tangent to $\Delta$, so that no projector
is needed. For a NEH, remarkably, this is in fact the case: Given $X$ and $Y$ tangent to $\Delta$, we have
\be
Y^{b}\nabla_{b}(X^{a}\ell_{a})
&\hateq& Y^{b}X^{a}\nabla_{b}\ell_{a} + Y^{b}\ell_{a}\nabla_{b}X^{a}\\
&\hateq& Y^{b}X^{a} \underleftarrow{\nabla_{b}\ell_{a}} + \ell_{a}Y^{b}\nabla_{b}X^{a} \; .
\ee
On $\Delta$, however, $X^{a}\ell_{a}=0$.  In addition, $\underleftarrow{\nabla_{(a}\ell_{b)}}\hateq0$ and $\ell$ is twist-free
so that $\underleftarrow{\nabla_{a}\ell_{b}}\hateq0$. Thus $\ell_{a}Y^{b}\nabla_{b}X^{a} \hateq 0$, so that $Y^{b}\nabla_{b}X^{a}$ is tangent to $\Delta$, as required.
It follows that $\nabla_a$ induces a well-defined derivative operator on $\Delta$, which we denote by $D_a$.
Because $\nabla_a$ is metric ($\nabla_a g_{bc} = 0$) and torsion-free, one deduces that $D_a$ is also metric ($D_a q_{bc} = 0$)
and torsion-free.  However, because $q_{ab}$ is degenerate, these two conditions
do \textit{not} uniquely determine $D_a$.  Thus, in contrast to the situation on space-like or time-like hypersurfaces, the derivative operator $D_a$ contains \textit{more information} than is contained in $q_{ab}$.

On a NEH, one can show that the vanishing of the expansion, shear and twist of $\ell^a$ implies that certain components of $D_a$
are reduced to a single intrinsic one-form $\omega_a$,
known as the induced normal connection:
\bea
\nabla_{\underleftarrow{a}}\ell_{b} \hateq D_{a}\ell_{b} \hateq \omega_{a}\ell_{b} \; .
\label{connectionondelta}
\eea
This quantity is gauge-dependent under rescalings of the null normal on $\Delta$.
Under transformations $\ell\rightarrow\ell^{\prime}=f\ell$ with $f$ some smooth function,
we find that
\be
D_{a}\ell_{b}^{\prime} &\hateq& fD_{a}\ell_{b} + (D_{a}f)\ell_{b}\\
                       &\hateq& \omega_{a}\ell_{b}^{\prime} + \frac{(D_{a}f)}{f}\ell_{b}^\prime \\
                       &\hateq& (\omega_{a} + D_{a}\ln f)\ell_{b}^{\prime} \; .
\ee
The induced normal connection $\omega_a^\prime$ associated to $\ell_a^\prime$ is thus
\bea
\omega_{a}^{\prime} = \omega_{a} + D_{a} \ln f \; .
\label{omegatransformation}
\eea
From (\ref{connectionondelta}), an expression for the surface gravity $\kappa_{(\ell)}$
can be isolated.  Contracting both sides of equation (\ref{connectionondelta}) with $\ell^{a}$,
\bea
\ell^{a}\nabla_{a}\ell^{b} = (\ell^{a}\omega_{a})\ell^{b} \, .
\eea
This equation is the geodesic equation for $\ell$ with non-zero acceleration.  Therefore the
surface gravity of $\Delta$ is
\bea
\kappa_{(\ell)} = \ell^{a}\omega_{a} \; \label{ihkappa}.
\eea
This quantity is likewise gauge-dependent under
rescalings of $\ell$.  For $\ell' = f \ell$,  $\kappa_{(\ell)}$ transforms as
\bea
\kappa_{(\ell^{\prime})} =f\ell^{a}(\omega_{a}+ D_{a} \ln f)=f\kappa_{(\ell)} + \pounds_\ell f .
\label{ktrans}
\eea
However, from (\ref{omegatransformation}), one sees that the curvature $d \omega$ of $\omega$ is
\textit{gauge-invariant}. In fact one can find an explicit expression for it.
We have the following.\\
\noindent{\bf Proposition 1.}
\emph{The curvature of the induced normal connection $\omega$ on $\Delta$ is given by}
\bea
d\omega \hateq 2\left(\mbox{Im}[\Psi_{2}]\right)\areaform \, ,
\label{curvatureofomega}
%
%
\eea
\emph{with} $\Psi_{2}$ \emph{the second Weyl scalar,} and $\areaform = i m \wedge \overline{m}$ the area element on $\Delta$.
%
%

\noindent{\bf Proof.}
We follow the proof that is presented in \cite{afk2000}.
Recall the definition of curvature, $2\nabla_{[a}\nabla_{b]}X^{c}=-R_{abd}^{\phantom{aaa}c}X^{d}$.
Applied to the case $X^{a}=\ell^a$, one has
$2\nabla_{[a}\nabla_{b]}\ell^{c}=-R_{abd}^{\phantom{aaa}c}\ell^{d}$.
Pulling back the $a,b$
indices and using
$\nabla_{\! \underleftarrow{a}}\ell_{b}\hateq\omega_{a}\ell_{b}$
from (\ref{connectionondelta}),
one obtains
\begin{equation}
2\ell^c D_{[a}\omega_{b]}\hateq-R_{\underleftarrow{ab}d}^{\phantom{aaa}c}\ell^{d}.
\label{curvmid}
\end{equation}
Furthermore, from (\ref{decomposition}) one has
$R_{\underleftarrow{ab}d}^{\phantom{aaa}c}\ell^{d}\hateq C_{\underleftarrow{ab}d}^{\phantom{aaa}c}\ell^{d}$ because $R_{\underleftarrow{ab}}\ell^{b}\hateq0$.  Combining this with (\ref{curvmid})
and contracting with $n_{c}$ gives $2 D_{[a}\omega_{b]}\hateq C_{\underleftarrow{ab}cd}\ell^{c}n^{d}$.
The Weyl tensor can be
expanded in terms of the complex scalars $\{\Psi\}$ as
\bea
C_{abcd}\ell^{c}n^{d}
&\hateq& 4\left(\mbox{Re}[\Psi_{2}]\right)n_{[a}\ell_{b]}
         + 2\Psi_{3}\ell_{[a}m_{b]} + 2\bar{\Psi}_{3}\ell_{[a}\bar{m}_{b]}\nonumber\\
& &      - 2\Psi_{1}n_{[a}\bar{m}_{b]} - 2\bar{\Psi}_{1}n_{[a}m_{b]}
         + 4\mbox{Im}[\Psi_{2}]m_{[a}\bar{m}_{b]} \; .
\eea
Pulling back the $a,b$ indices and substituting in the curvature, the expression (\ref{curvatureofomega})
follows.  \qedd

\subsection{Isolated horizons}

It is clear from the transformation law (\ref{ktrans}) for surface gravity that, on a given NEH, the zeroth law of black hole mechanics cannot hold for all possible null normals.
We now ask the question: for which, if any, null normals does it hold?
The Cartan identity reads
\be
\pounds_{\ell}\omega_{b} \hateq
2\ell^{a}D_{[a}\omega_{b]} + D_{b}(\ell^{a}\omega_{a}) \; .
\ee
Combining this with (\ref{curvatureofomega}), we have
\bea
0 \hateq 4 \ell^a \mbox{Im}[\Psi_{2}]m_{[a}\bar{m}_{b]} \hateq \pounds_{\ell}\omega_{b} - D_{b}(\ell^{a}\omega_{a})  \, ,
\eea
%
%
so that $D_{b}\kappa_{(\ell)}\hateq\pounds_{\ell}\omega_{b}$.
It follows that the surface gravity is constant over the entire NEH \textit{iff} $\pounds_{\ell}\omega_{b}\hateq0$, i.e., iff
$\omega_a$ is in `equilibrium' with respect to $\ell$.

The condition $\pounds_\ell \omega_a \hateq 0$ can also be interpreted in terms of
`extrinsic curvature'.
Strictly speaking, because $\Delta$ is null, it does not have an extrinsic curvature in the usual sense.
Nevertheless, one can define an analogue of extrinsic curvature by using the same formula that is used for space-like surfaces involving the
Levi-Civita derivative operator and normal to the surface:
\bea
K_{a}{}^b \equiv \nabla_{\! \underleftarrow{a}}\ell^{b} \hateq D_{a}\ell^{b} \; .
\label{KDelta}
\eea
Note that, because $\Delta$ is a null surface, this analogue of extrinsic curvature has the curious property that it is
fully determined by the \textit{intrinsic} structures $D_a$ and $\ell_a$, so that the nomenclature ``extrinsic geometry''
is only appropriate by analogy, and should not be taken too literally. From (\ref{KDelta}), one sees that,
on a NEH, fixing the extrinsic geometry of $\Delta$ to be time independent is equivalent to fixing the induced
normal connection $\omega_a$ of $\Delta$ to be time independent, which in turn, as we saw above, is the necessary and sufficient
condition for the (gravitational) zeroth law to hold.

Note furthermore that if this condition holds for a single null normal $\ell$, then it will hold
for all other null normals $\ell' = c \ell$ related by a \textit{constant} rescaling.
That is, if we wish the zeroth law to hold, it is sufficient to restrict to
an \textit{equivalence class} of null normals, where two normals are equivalent if they are related by a constant rescaling.
Note the similarity to Killing horizons, where one has
the freedom to rescale the Killing field $\xi^a$ only by a \textit{constant}.

In Einstein-Maxwell theory, one can establish a similar zeroth law for the electric potential
$\Phi{(\ell)} := \ell^a \bm{A}_a$ as follows.
First, the energy condition imposed in
Definition 1 implies that the Maxwell field satisfies $\underleftarrow{\ell\lrcorner\bm{F}}\hateq0$.\footnote{Here and throughout this chapter
$\cdot \lrcorner \cdot$ denotes contraction of a vector with the first index of a form.}
Using this, together with the Cartan identity
and the Bianchi identity ($d\bm{F}=0$), this implies
$\pounds_{\ell}\underleftarrow{\bm{F}}=d(\underleftarrow{\ell\lrcorner\bm{F}}) + \underleftarrow{\ell\lrcorner
d\bm{F}}\hateq0$.  It follows that the electromagnetic potential $\bm{A}$ can be partially gauge-fixed such
that $\pounds_{\ell}\underleftarrow{\bm{A}}\hateq0$.  This is referred to as a \textit{gauge adapted to the horizon}.
When this is satisfied, it follows that
$0=\pounds_{\ell}\underleftarrow{\bm{A}}=d(\underleftarrow{\ell\lrcorner\bm{A}})
+\underleftarrow{\ell \lrcorner d\bm{A}}=d\underleftarrow{\Phi_{(\ell)}}$, where we used the
condition $\underleftarrow{\ell\lrcorner\bm{F}}\hateq0$.  It follows that the electric potential is constant
over the entire NEH, so that the zeroth law also holds for the electric potential
$\Phi{(\ell)}$. The above observations lead to the following definition.

\noindent{\bf Definition 2. (Weakly Isolated Horizon).}
\emph{A NEH $\Delta$ equipped with an equivalence class $[\ell]$ of future-directed null normals, with $\ell^\prime \sim \ell$
if $\ell^\prime =c\ell$ ($c>0$ a constant), such that the  $\pounds_{\ell}\omega_{a}\hateq0$ and
$\pounds_{\ell}\pb{\bm{A}}=0$ for all $\ell\in[\ell]$, is said to be a weakly isolated horizon (WIH).}

Note that because $c$ is now a constant, $\omega_{a}$ is uniquely determined by the equivalence class $[\ell]$.  Under
the re-scaling $\ell^{\prime}=c\ell$, the surface gravity $\kappa_{(\ell)}$ transforms as
$\kappa_{\ell^{\prime}}=c\kappa_{(\ell)}$. However, the condition $\kappa_{(\ell)}=0$ is independent of the choice
of $\ell \in [\ell]$.  If $\kappa_{(\ell)} = 0$, we say
the WIH is \textit{extremal}, in analogy with the nomenclature for Killing horizons.
When discussing a WIH, it is convenient at this point to strengthen the partial gauge-fixing of the tetrad so that
$\ell^a \in [\ell]$, and additionally $d n = 0$.
%
%
Via the Cartan identity, the latter condition implies $\pounds_\ell n_a = 0$.

Let us now ask: Given a NEH, under what conditions does there exist an equivalence class $[\ell]$
of null normals such that it becomes a WIH?  The answer to this question turns out to be \textit{always}
\cite{abl2001a}.  Thus, as far as geometry is concerned, the WIH definition is not more restrictive
than the NEH definition.  Rather, the importance of the WIH definition, as compared to the NEH definition,
lies in the selection of an equivalence class $[\ell]$.
%
%

There is also a stronger notion of isolated horizon that one can introduce.
On a WIH, while the normal connection $\omega_a$ on a WIH is time independent, the other components of the connection $D_a$
can in general vary.  Thus, a stronger condition would be to impose that the \textit{entire} connection $D_a$ on $\Delta$
possess $\ell^a$ as a symmetry.  This condition is equivalent to
$[\pounds_{\ell},D_a]=0$.  By contrast, the condition $\pounds_\ell \omega_a = 0$ in Definition 2 is equivalent to the weaker
condition $[\pounds_{\ell},D_a]\ell^{b}=0$.
%
%
%
We have the following.

\noindent{\bf Definition 3. (Strongly Isolated Horizon).}
\emph{A NEH $\Delta$ equipped with an equivalence class $[\ell]$ of future-directed null normals such that
$[\pounds_{\ell},D_{a}]X^{b}\hateq 0$ for \emph{every} $X^{b}$ tangent to $\Delta$ is said to be
a strongly isolated horizon (SIH).}

In particular, every Killing horizon is also a SIH because all of the geometry at $\Delta$ (including the connection) possesses $\ell^a$ as a symmetry.

Given a NEH,
the above stronger condition cannot always be met by simply choosing the equivalence class $[\ell]$ appropriately -- to be a SIH involves
a genuinely stronger restriction on the geometry of the horizon.  When an equivalence class $[\ell]$ does exist making an NEH into a SIH,
it is furthermore \textit{unique} \cite{abl2001a}.  The sole remaining ambiguity in the choice of $\ell^a$ is then to rescale it by a constant, similar to the ambiguity to rescale by a constant the Killing vector field in a stationary space-time.  In the case of a stationary space-time, this rescaling freedom can be eliminated by requiring the Killing vector field to be unit at spatial infinity.
As we shall see in the next section, for a SIH in the context of Einstein-Maxwell theory,
one can similarly remove the remaining constant rescaling ambiguity in $\ell^a$, but this time
in a way that is purely \textit{quasi-local}.  For the moment, we
leave the freedom intact.

For most of the rest of this chapter, the results we consider
will require only weakly isolated horizons. For this reason, hence forth, the term `isolated horizon'(IH), when not
otherwise qualified, shall refer specifically to a \textit{weakly} isolated horizon.

\subsection{The topology of strongly isolated horizons}
\label{topsect}

Let us make a couple of further definitions.
If the expansion of $n^a$ is negative on $\Delta$, $\theta_n := q^{ab} \nabla_a n_b < 0$, let us call $\Delta$
a \textit{future isolated horizon}.  In this case $\Delta$ describes a black hole horizon and not a white hole horizon.
Furthermore, recall that, although $\kappa_{(\ell)}$ in general depends on the choice of $\ell$ in the equivalence class $[\ell]$,
the \textit{sign} of $\kappa_{(\ell)}$ does not. We have already defined $\Delta$ to be \textit{extremal} if
$\kappa_{(\ell)} = 0$.  Furthermore, for future isolated horizons, if $\kappa_{(\ell)} > 0$, we call $\Delta$ \textit{sub-extremal}, and if $\kappa_{(\ell)} < 0$
we call $\Delta$ \textit{super-extremal}.
We then have the following result:

\noindent{\bf Proposition 2.}
\emph{Suppose $\Delta$ is a future SIH in a space-time with zero cosmological constant.  If $\Delta$ is sub-extremal, then
its cross-sections have 2-sphere topology.  If $\Delta$ is extremal, then its cross-sections can have either 2-sphere
or 2-torus topology, with the 2-torus topology occurring iff both $\tilde{\omega}=0$ and
$T_{ab}\ell^{a}n^{b}\hateq 0$.}

\noindent{\bf Proof.}

Consider the evolution equation for the expansion of the auxiliary null normal $n^{a}$
(see \cite{bf2007a}, or equivalently (3.7) in \cite{abl2001a}):
%
%
\bea
\pounds_{\ell}\theta_{(n)} + \kappa \theta_{(n)} + \frac{1}{2}\mathcal{R}
= \slD_{a}\tilde{\omega}^{a} + \|\tilde{\omega}\|^{2} +  (\Lambda - 8\pi G T_{ab} \ell^{a}n^{b}) \; .
\label{nevolution}
\eea
Here, $\mathcal{R}$ is the scalar curvature of $\mathcal{S}$, $\slD_{a}$ is the covariant derivative operator that
is compatible with the metric $\tilde{q}_{ab}=g_{ab}+\ell_{a}n_{b}+\ell_{b}n_{a}$ on $\IHslice$, and
$\|\tilde{\omega}\|^{2}=\tilde{\omega}_{a}\tilde{\omega}^{a}$ with
$\tilde{\omega}_{a}=\tilde{q}_{a}^{\phantom{a}b}\omega_{b}=\omega_{a}+\kappa_{(\ell)}n_{a}$ the projection of $\omega$
onto $\IHslice$. From strong isolation, and $\pounds_\ell n = 0$, one has
$\pounds_\ell \theta_{(n)} = \pounds_\ell q^{ab} D_a n_b = q^{ab} D_a \pounds_\ell n_b = 0$.
Using this fact with $\Lambda = 0$, and $\theta_{(n)}<0$, (\ref{nevolution}) becomes
\bea
- \kappa |\theta_{(n)}| = -\frac{1}{2}\mathcal{R} + \slD_{a}\tilde{\omega}^{a} + \|\tilde{\omega}\|^{2}
                     + T_{ab} \ell^{a}n^{b} \; .
\label{nevolution2}
\eea
Integrating both sides over the surface $\IHslice$,
and finally using that $\oint_{\IHslice}\areaform \mathcal{D}_{a}\tilde{\omega}^{a}=0$, one has
\bea
\oint_{\IHslice}\areaform \mathcal{R}
\geq 2\oint_{\IHslice}\areaform(T_{ab}\ell^{a}n^{b} + \kappa |\theta_{(n)}| + \|\tilde{\omega}\|^{2}) \; .
\label{curvaturecondition}
\eea
The dominant energy condition requires that $T_{ab}\ell^{a}n^{b}\geq0$.  In addition, because we have excluded the super-extremal case, the second term is non-negative. Lastly, $\|\tilde{\omega}\|^{2}$ is
manifestly non-negative.  It follows that $\oint_\IHslice \areaform \mathcal{R} \ge 0$ with equality iff $\kappa_{(\ell)}=0$, $T_{ab} \ell^a n^b = 0$
and $\tilde{\omega} = 0$.
On the other hand, the Gauss-Bonnet theorem gives
\begin{displaymath}
\oint_\IHslice \areaform \mathcal{R} = 8\pi(1-g)
\end{displaymath}
%
%
where $g$ is the genus of $\IHslice$, so that $\oint_\IHslice \areaform \mathcal{R} > 0$ iff $\IHslice$ is a 2-sphere,
and $\oint_\IHslice \areaform \mathcal{R} = 0$ iff $\IHslice$ is a 2-torus.  Thus, if $\Delta$ is sub-extremal
($\kappa_{(\ell)} > 0$), one has $\oint_\IHslice \areaform \mathcal{R} > 0$, so that $\IHslice$ is a 2-sphere,
whereas if $\Delta$ is extremal, both 2-sphere and 2-torus topologies are possible, with the latter occurring
iff $T_{ab} \ell^a n^b = 0$ and $\tilde{\omega} = 0$.
\qedd

%
%

\subsection{Existence of Killing spinors}

The extremal Kerr-Newman black hole is a solution to the $N=2$ supergravity field equations
with the fermion fields set to zero.  The condition for this solution to have positive energy
is that \cite{gh1982}
\bea
\mathfrak{M}=|\mathfrak{Q}| \, ,
\label{bps2}
\eea
relating the mass $\mathfrak{M}$ and total charge $\mathfrak{Q}\equiv\sqrt{q_{e}^{2}+q_{m}^{2}}$ (with $q_{e}$
and $q_{m}$ the electric and magnetic charges); this is the extremality condition for the Kerr-Newman black
hole.  This is also the saturated Bogomol'ny-Prasad-Sommerfeld (BPS) inequality \cite{gh1982}.
This leads to an interesting question: If we try to define a supersymmetric \textit{isolated horizon},
do we have a similar restriction to extremality?

Generally, Einstein-Maxwell theory (in four dimensions) can be viewed as the bosonic sector of $N=2$
supergravity --- that is, the sector in which the spin-$3/2$ gravitino field and its complex conjugate
vanish in vacuum.  In this sector, the defining property of supersymmetric configurations is the
existence of a \emph{Killing spinor}, whence we look for a restricted space of solutions having this
property.
In the presence of zero cosmological constant, a Killing spinor can be defined as a (Dirac) spinor
$\zeta$ satisfying
\bea
\left[\nabla_{\! a} + \frac{i}{4}\bm{F}_{bc}\gamma^{bc}\gamma_{a}\right]\zeta = 0 \; .
\label{cond1}
\eea
Here, $\gamma^{a}$ are a set of gamma matrices that satisfy the usual anticommutation rule
\bea
\gamma^{a}\gamma^{b} + \gamma^{b}\gamma^{a} = 2g^{ab}
\eea
and the antisymmetry product
\bea
\gamma_{abcd} = \epsilon_{abcd} \; .
\eea
%
%
$\gamma_{a_{1}\ldots a_{D}}$ denotes the antisymmetrized product of $D$ gamma matrices.  The complex
conjugate $\bar{\zeta}$ of $\zeta$ is defined as
\bea
\bar{\zeta} = i(\zeta)^{\dagger}\gamma_{0} \, ;
\eea
with $\dagger$ denoting Hermitian conjugation.

From $\zeta$ and $\bar{\zeta}$ one can construct five (real) bosonic bilinear covariants
\bea
f = \bar{\zeta}\zeta \, ,
\quad
g = i\bar{\zeta}\gamma^{5}\zeta \, ,
\quad
V^{a} = \bar{\zeta}\gamma^{a}\zeta \, ,
\quad
W^{a} = i\bar{\zeta}\gamma^{5}\gamma^{a}\zeta \, ,
\quad
\Psi^{ab} = \bar{\zeta}\gamma^{ab}\zeta \; .
\eea
These bilinear covariants are all related to each other via several algebraic conditions (from the
Fierz identity) and differential equations (from the Killing spinor equation (\ref{cond1})).  In
particular, the vector $V$ satisfies the equations
\bea
V_{a}V^{a} &=& -(f^{2} + g^{2}) \, ,\label{tnvector1}\\
\nabla_{a}V_{b} &=& - f\bm{F}_{ab} + \frac{g}{2} \epsilon_{abcd}\bm{F}^{cd} \; .
\label{Psi1}
\eea
%

The above reviews the standard notion of a Killing spinor.  What we now wish to consider
is a Killing spinor which exists
\textit{on the horizon $\IH$ only} (which we assume to be strongly isolated).
Then, in place of (\ref{cond1}), we can at most impose
a version with the derivative pulled back to $\IH$:
\begin{equation}
\left[\nabla_{\! \pb{a}} + \frac{i}{4}\bm{F}_{bc}\gamma^{bc}\gamma_{\pb{a}}\right]\zeta = 0 \; .
\label{pbcond}
\end{equation}
which, in place of (\ref{Psi1}), leads to
\begin{displaymath}
\nabla_{\pb{a}}V_{b} = - f\bm{F}_{\pb{a}b} + \frac{g}{2} \epsilon_{\pb{a}bcd}\bm{F}^{cd}.
\end{displaymath}
If one additionally pulls back $b$ to $\IH$ and symmetrizes, one sees that
\begin{displaymath}
\pounds_V q_{ab} = 2 \pb{\nabla_{(a} V_{b)}} = 0 ,
\end{displaymath}
which is the same as the condition (\ref{qsymm}) satisfied by $\ell$ on $\Delta$ as a NEH.
If we furthermore stipulate that $V$ be \textit{equal} to a null normal
in the equivalence class $[\ell]$ making $\Delta$ a SIH, we have the notion of a \textit{supersymmetric isolated horizon}.

\noindent{\bf Definition 4. (Supersymmetric Isolated Horizon).}
\emph{A SIH $\Delta$ equipped with a Killing spinor $\zeta$ and complex conjugate
$\bar{\zeta}=i(\zeta)^{\dagger}\gamma_{0}$,
such that $V:=\bar{\zeta}\gamma^{a}\zeta\in[\ell]$, is said to be a supersymmetric isolated horizon (SSIH).}

For SSIHs, the following can be proved.

\noindent{\bf Proposition 3.}
\emph{An SSIH of Einstein-Maxwell theory with zero cosmological constant is necessarily extremal
and `non-rotating' --- that is, has zero gravitational angular momentum as defined in equation (\ref{closedforms}) below.}

\noindent{\bf Proof.}
For a SSIH, $V^{a}$ and $\ell^{a}$ are identified.
Hence, at $\IH$, $V^a$ is null so that
from (\ref{tnvector1}) we have $f=g=0$.  Equation
(\ref{connectionondelta}) together with (\ref{Psi1}) then gives
\bea
\nabla_{\underleftarrow{a}} \ell_{b} =  \omega_a \ell_b = 0 \, ,
\label{psipullback}
\eea
so that $\omega\hateq0$.  This condition implies that the gravitational angular momentum, defined in equation (\ref{closedforms})
below, is identically zero.  However, in general $\bm{A}$ is non-zero, which means
that there may be non-zero angular momentum stored in the
%
%
electromagnetic fields.  The condition
also implies that $\kappa_{(\ell)}=\ell\lrcorner\omega=0$.  Therefore, \emph{SSIHs are extremal and non-rotating}.
\qedd

\section{Hamiltonian mechanics}

Up until now we have been considering geometric properties of isolated horizons.  At this point
we show how the isolated horizon boundary conditions lead to a \textit{consistent variational and canonical framework}. Specifically, we start with an action principle, derive from this the covariant phase space, and then finally discuss the definition of angular momentum and mass of isolated horizons as generators of rotations and time translations.  This will lead directly to a derivation of the first law of black hole mechanics involving only quantities which are \textit{quasi-locally defined} in terms of fields intrinsic to the horizon.

%
%
For conceptual clarity, in this presentation we focus on
Einstein-Maxwell theory. Furthermore, we use the Palatini formulation of gravity in terms of a co-tetrad $e_a^I$ and
associated internal Lorentz connection $A_a^{IJ}$, where $I,J = 0,1,2,3$ are internal indices.
This will greatly simplify the necessary formulae,
and allow an easier transition to the brief discussion on quantum theory at the end of the chapter.
In this formulation, the space-time metric is constructed from the co-tetrad as $g_{ab} = \eta_{IJ}e_a^I e_b^J$ where
$\eta_{IJ} = {\rm diag}(-1,1,1,1)$ is the internal Lorentz metric, and the curvature $F^{IJ}_{ab}$ of $A_a^{IJ}$ determines
the Riemann tensor via $F^{IJ}_{ab} = \left(d A^{IJ} + A^I{}_K \wedge A^K{}_J\right)_{ab} = R_{abc}{}^d e^{cI} e_d^J$.
Internal indices are raised and lowered with $\eta^{IJ}, \eta_{IJ}$.
We denote the Maxwell vector potential
by $\bm{A}$, so that the field strength is $\bm{F} = d \bm{A}$.

\subsection{Action principle}

\begin{figure}[t]
\begin{center}
\psfrag{D}{$\Delta$}
\psfrag{I}{$\mathscr{I}$}
\psfrag{Mp}{$M_{2}$}
\psfrag{Mm}{$M_{1}$}
\psfrag{Mi}{$M$}
\psfrag{Sp}{$\phantom{Sp}$}
\psfrag{Sm}{$\phantom{Sm}$}
\psfrag{S}{$\IHslice$}
\psfrag{I}{$\mathcal{I}$}
\psfrag{C}{$\IHslice_\infty$}
\psfrag{M}{$\mathcal{M}$}
\includegraphics[width=4.8in]{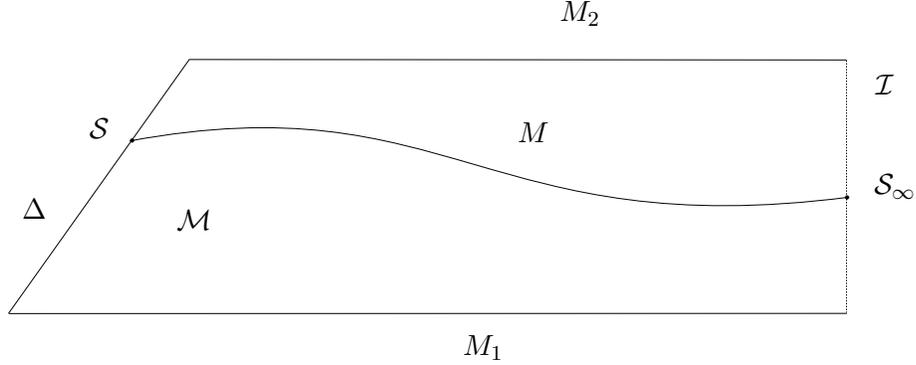}
\caption{
The region of space-time $\mathcal{M}$ considered has an internal boundary $\Delta$
which will be the isolated horizon, and is bounded to the past and future by two three-dimensional Cauchy surfaces $M_{1}$
and $M_{2}$.
$M$ is a partial Cauchy surface that intersects $\Delta$ and $\mathcal{I}$ each in a two-sphere.
\label{arena}}
\end{center}
\end{figure}

Let us consider the action for Einstein-Maxwell theory in the Palatini formulation
on a four-dimensional manifold $\mathcal{M}$ with boundary
$\partial\mathcal{M} \cong M_{1} \cup M_{2} \cup \Delta \cup \mathcal{I}$.
Here $\mathcal{I}$ represents the two-sphere at spatial infinity crossed with time, where appropriate asymptotically
flat boundary conditions are imposed.
$\Delta$ is a three-dimensional manifold with topology $S^2 \times [0,1]$
constrained to be an isolated horizon equipped with a fixed equivalence class of null normals $[\ell_a]$,
with all equations of motion holding at $\Delta$.
$M_1, M_2$ play the role of partial Cauchy surfaces.
As before, the isolated horizon boundary conditions are understood to include the requirement
that the Maxwell potential be in a gauge adapted to the horizon, $\pounds_\ell \pb{\bm{A}} = 0$.
The space-time region $\mathcal{M}$ thus described is shown in Figure \ref{arena}.

The action is then given by
%
%
\bea
S &=& \frac{1}{16\pi G}\int_{\mathcal{M}}\Sigma_{IJ} \wedge F^{IJ}
      - \frac{1}{16\pi G}\int_{\mathcal{I}}\Sigma_{IJ} \wedge A^{IJ}
      - \frac{1}{8\pi}\int_{\mathcal{M}}\bm{F} \wedge \star \bm{F}  \; .
\label{action}
\eea
where  $\bm{F}=d\bm{A}$ is the field strength of the Maxwell field, $\star$ denotes Hodge dual, and
$\Sigma^{IJ}:= \frac{1}{2} \epsilon^{IJ}{}_{KL} e^K \wedge e^L$ with $\epsilon_{IJKL}$ the internal alternating tensor.
The boundary term at $\mathcal{I}$
is necessary to make the action differentiable.  By contrast, as we will show below, \textit{no} term at $\Delta$
is necessary for the action to be differentiable \cite{aes2008,as2008,ls2008}.

Let us denote the dynamical variables $(e,A,\bm{A})$ collectively by $\Psi$.  Application of an arbitrary variation
$\delta$ to the action then yields the form
%
%
\bea
\delta S = \int_{\mathcal{M}}E[\Psi] \cdot \delta\Psi - \int_{\partial\mathcal{M}}\theta(\delta)[\Psi] \; .
\label{first}
\eea
Here $E[\Psi]=0$ denotes the equations of motion.  Specifically, these are:
\bea
&& \epsilon_{IJKL}e^{J} \wedge F^{KL} = T_{I} \, ,
\label{em1}\\
&& \d\Sigma_{IJ} - A_{I}{}^{K} \wedge \Sigma_{KJ} - A_{J}{}^{K} \wedge \Sigma_{IK} = 0 \, ,
\label{em2}\\
&& \d \star \bm{F} = 0 \; .
\label{em3}
\eea
where in the first of these equations, $T_{I}$ denotes the electromagnetic stress-energy three-form.
The second equation imposes that $A^{IJ}_a$ be the unique torsion-free connection
compatible with $e_a^I$. Together the above equations are equivalent to the Einstein-Maxwell
equations in metric variables, with the components of $T_{I}$ identified with appropriate components the
electromagnetic stress-energy tensor.

The integrand $\theta(\delta)$ of the surface term in (\ref{first}) is given by
\bea
\theta(\delta) = \theta_{\rm Grav}(\delta) + \theta_{\rm EM}(\theta) \, ,
\label{surface}
\eea
where we have defined
$\theta_{\rm Grav}(\delta) := (1/16\pi G)\Sigma_{IJ} \wedge \delta A^{IJ}$
and
$\theta_{\rm EM}(\delta) := (1/16\pi G)\star \bm{F} \wedge \delta \bm{A}$,
and where we suppress explicit representation of the dependence on
$\Psi = (e, A, \bm{A})$.

The action $S$ is said to be differentiable if, when the configuration fields
$e, A, \bm{A}$ are fixed on $M_1$ and $M_2$, the boundary term in (\ref{first}) vanishes.
Let us show that this is the case.  Because $A$ and $\bm{A}$ are held fixed at $M_{1},M_{2}$, $\theta(\delta)$
vanishes there.  In addition, the boundary term in the action (\ref{action}) is constructed
precisely such that the boundary terms at $\mathcal{I}$
coming from the variation of the action cancel.  Therefore it remains only
to show that the integral of $\theta(\delta)$ over $\Delta$ vanishes.

To see that this is true, let us define an \textit{internal} Newman-Penrose basis $\ell^I =(1,1,0,0)/\sqrt{2}$,
$n^I =(1,-1,0,0)/\sqrt{2}$, $m^I =(0,0,1,i)/\sqrt{2}$, $\overline{m} =(0,0,1,-i)/\sqrt{2}$.  The relations
(\ref{npdef}) then become
\begin{displaymath}
\ell^a = e^a_I \ell^I, \quad n^a = e^a_I n^I, \quad m^a = e^a_I m^I, \quad \overline{m}^a = e^a_I \overline{m}^I.
\end{displaymath}
From the definition of $\omega$ (\ref{connectionondelta}), one then deduces the equation
\be
A_{\underleftarrow{a}IJ}\ell^{J}\hateq\omega_{a}\ell_{I} \, .
\ee
By expanding the pull-backs to $\Delta$ of $A^{IJ}$ and $\Sigma^{IJ}$ in terms of the Newman-Penrose
basis, using the above identity, and simplifying, one obtains
\begin{displaymath}
\underleftarrow{\Sigma}_{IJ} \wedge \delta\underleftarrow{A}^{IJ}\hateq 2\areaform \wedge
\delta\omega \, ,
\end{displaymath}
with $\areaform=m\wedge\overline{m}$ the area two-form on $\Delta$.  One thus has
\bea
\underleftarrow{\theta}_{\rm Grav}(\delta) \hateq \frac{1}{16\pi G}\areaform \wedge \delta\omega \; .
\label{pullbackofcurrent}
\eea
Throughout the space of histories considered, $\pounds_\ell \omega = 0$, so that $\pounds_\ell \delta \omega = 0$.
This, combined with the fact that $A^{IJ}_a$, $e^I_a$, and hence $\omega$, is fixed on
$\IHslice_1 := M_1 \cap \Delta$ and $\IHslice_2:= M_2 \cap \Delta$,
implies $\delta \omega = 0$ on all of $\Delta$, so that $\pb{\theta}_{\rm Grav}(\delta)=0$.  The argument for
the electromagnetic part $\pb{\theta}_{\rm EM}(\delta)$ is similar: Because $\bm{A}$ is in a
gauge adapted to the horizon, so that $\pounds_\ell \pb{\bm{A}} = 0$ throughout the space of histories, we have
$\pounds_\ell \delta \pb{\bm{A}} = 0$.  This combined with the fact that $\bm{A}$ is fixed on $\IHslice_1$ and $\IHslice_2$
implies $\delta \pb{\bm{A}} = 0$, implying $\pb{\theta}_{\rm EM}(\delta) = 0$.
This shows that, when the configuration variables $(e, A, \bm{A})$ are fixed on $M_1$ and $M_2$, the boundary term
in the variation of the action (\ref{first}) vanishes, so that $\delta S = 0$ implies the equations of
motion, as required.

\subsection{Covariant phase space}

The covariant phase space of a theory is the space of solutions of the equations
of motion.  Because it is not formulated in terms of initial data,
such a formulation of phase space enables a \textit{space-time covariant} version of
the canonical framework.  Furthermore, in this approach, one can derive the symplectic
structure directly through (anti-symmetrized) second variations of the
action, as we shall review. The notion of the covariant phase space as the space of
solutions, and the corresponding method of determining the symplectic structure
can be traced back to the work of Lagrange himself (see \cite{abr1991, lw1990}).

The covariant phase space $\bm{\Gamma}_{\rm Cov}$ and its symplectic structure $\Omega_{\rm Cov}$
are directly related to the more standard canonical phase space $\bm{\Gamma}$ and its
symplectic structure $\Omega$, through the fact that the space of initial data is
in one-to-one correspondence with the space of solutions.  This isomorphism furthermore
maps the covariant symplectic structure into the canonical symplectic structure, so that the
two frameworks are completely equivalent.

In the present context, the situation is more subtle, for two reasons: First, general relativity
is a \textit{constrained} theory, and usually one works with the phase space of unconstrained
initial data.  By contrast, in the covariant phase space, all equations of motion are by definition
satisfied. Second, the space-times under consideration are not globally hyperbolic, but rather only admit
partial Cauchy surfaces, with inner boundary at the isolated horizon $\Delta$. Consequently,
solutions in the covariant phase space have more information than is present in the initial data
on any given spatial hypersurface --- solutions in the covariant phase space ``know'' what fell into the black
hole in the past, whereas the initial data does not.  Nevertheless, the two frameworks can be related.
If we let $\overline{\bm{\Gamma}}$
denote the space of constrained initial data, one has a projector
$\pi_M: \bm{\Gamma}_{\rm Cov} \rightarrow \overline{\bm{\Gamma}}$ mapping a given solution
to the initial data which it induces on a given fixed hypersurface $M$.  Using this projector, and the inclusion map
$\iota: \overline{\bm{\Gamma}} \hookrightarrow \bm{\Gamma}$, one can show that
\begin{displaymath}
(\iota \circ \pi_M)^* \bm{\Omega} = \bm{\Omega}_{\rm Cov}
\end{displaymath}
so that the symplectic structure on the usual unconstrained phase space of initial data
is exactly mapped into that on the covariant phase space computed using the second variations
of the action.

In the following we will use the covariant phase space framework.  This will not only make the
space-time geometry more transparent, but will also simplify many calculations. More specifically, let
${\bm \Gamma}_{\rm IH}$ denote \textit{the covariant phase space of possible fields $(e,A,\bm{A})$
on $\mathcal{M}$ (1.) satisfying the Einstein-Maxwell equations, (2.) possessing appropriate
asymptotically flat fall off conditions at infinity,
%
%
and (3.) such that $\Delta$ is an isolated horizon with fixed equivalence class of null normals $[\ell]$.}

\subsection*{Symplectic structure on ${\bm \Gamma}_{\rm IH}$}

\begin{figure}[t]
\begin{center}
\psfrag{D}{$\Delta$}
\psfrag{I}{$\mathscr{I}$}
\psfrag{Mp}{$M_{2}$}
\psfrag{Mm}{$M_{1}$}
\psfrag{Sp}{$\phantom{Sp}$}
\psfrag{Sm}{$\phantom{Sm}$}
\psfrag{I}{$\mathcal{I}$}
\psfrag{o}{$\bm{\omega}[\delta_{1},\delta_{2}]$}
\includegraphics[width=4.8in]{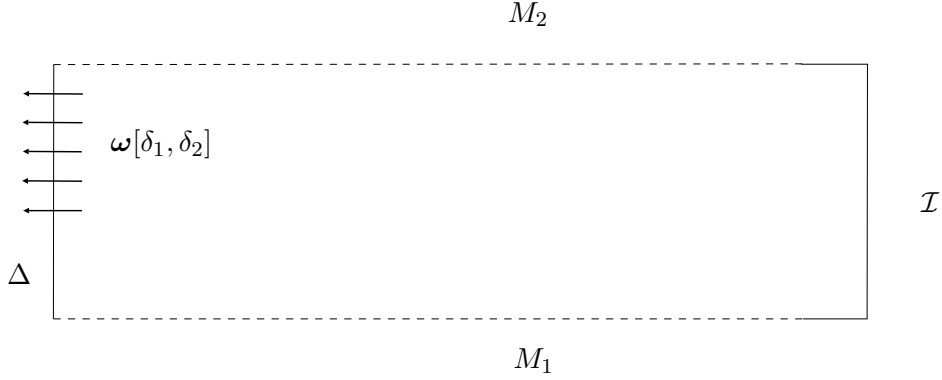}
\caption{Symplectic current escaping across the horizon.\label{syl_escape}}
\end{center}
\end{figure}

The boundary term $\theta(\delta)$ in the variation of the action (\ref{first})
also determines the symplectic structure.  Though $\theta(\delta)$ is a 3-form on
space-time, as it is linear function of a single variation $\delta$,
it is also a \textit{1-form on phase space}. One can therefore take its exterior
derivative to obtain a \textit{2-form} on phase space, called the
\textit{symplectic current}, $\bm{\omega} := \dbldif \bm{\theta}$,
where $\dbldif$ denotes the exterior derivative on (the infinite dimensional space)
$\bm{\Gamma}_{\rm IH}$.
Explicitly, one obtains
\bea
\bm{\omega}(\delta_1, \delta_2)
&=& \frac{1}{16 \pi G}
    \left[\delta_{1}A^{IJ} \wedge \delta_{2}\Sigma_{IJ}
    - \delta_{2}A^{IJ} \wedge \delta_{1}\Sigma_{IJ}\right]\nonumber\\
& & + \frac{1}{4\pi} \left[\delta_{1}\bm{A} \wedge \delta_{2}(\star \bm{F})
    - \delta_{2}\bm{A} \wedge \delta_{1}(\star \bm{F}) \right] \; .
\eea
The symplectic current has the property that for any two variations $\delta_1$, $\delta_2$ tangent to
the space of solutions, $\bm{\omega}(\delta_1, \delta_2)$ is closed:
$d \bm{\omega}(\delta_1, \delta_2) = 0$.
Normally one would then define the symplectic structure for a given Cauchy
hypersurface $M$ to be
\begin{displaymath}
\bm{\Omega}_M^B(\delta_1, \delta_2) :=\int_M \bm{\omega}(\delta_1, \delta_2) \; .
\end{displaymath}
However, in the present case, because of the presence of the inner boundary $\Delta$,
this symplectic structure is not conserved in time.  That is: for two different
partial Cauchy surfaces $M_1, M_2$, $\bm{\Omega}^B_{M_1} \neq \bm{\Omega}^B_{M_2}$.
The reason for this can be seen in figure \ref{syl_escape}: symplectic current
is escaping across the horizon $\Delta$.  More precisely, because the symplectic
current is closed ($d \bm{\omega}(\delta_1, \delta_2) = 0$), Stokes theorem
implies $\oint_{\partial \mathcal{M}} \bm{\omega}(\delta_1, \delta_2) = 0$, so that
\be
\int_{M_{2}}\bm{\omega}(\delta_{1},\delta_{2})
= \int_{M_{1}}\bm{\omega}(\delta_{1},\delta_{2}) - \int_{\Delta}\bm{\omega}(\delta_{1},\delta_{2}) \; .
\ee
(The part of the boundary integral at spatial infinity $\mathcal{I}$ vanishes due to the imposition
of fall-off conditions.)  The solution to this problem is to use the IH boundary conditions to rewrite
the integral over $\Delta$ as an integral over the two-sphere intersections $\IHslice_1$ and $\IHslice_2$ of $\Delta$
with $M_1$ and $M_2$:
\bea
\label{deltaflux}
\int_{\Delta}\bm{\omega}(\delta_{1},\delta_{2}) = \left(\oint_{\IHslice_{2}} - \oint_{\IHslice_{1}}\right)
                                                          \bm{\lambda}(\delta_{1},\delta_{2})
\eea
for some two-form $\bm{\lambda}(\delta_1, \delta_2)$ on $\Delta$.  If one then defines the full symplectic
structure on a spatial slice $M$ to be
\begin{displaymath}
\bm{\Omega}_M (\delta_2, \delta_1) = \int_M \bm{\omega}(\delta_1, \delta_2) + \oint_\IHslice \bm{\lambda}(\delta_1, \delta_2) \, ,
\end{displaymath}
with $\IHslice=M\cap\Delta$, then conservation of the symplectic structure is restored, giving
$\bm{\Omega}_{M_1}(\delta_2, \delta_2) = \bm{\Omega}_{M_2}(\delta_1, \delta_2)$.

To carry this out for the present case, one defines potentials
$\psi$ and $\chi$ for the surface gravity $\kappa_{(\ell)}$ and electric potential $\Phi_{(\ell)}$ such that
\bea
\pounds_{\ell}\psi \hateq \ell \lrcorner \omega = \kappa_{(\ell)}
\quad
\mbox{and}
\quad
\pounds_{\ell}\chi \hateq \ell \lrcorner \bm{A} = -\Phi_{(\ell)} \, .
\label{potentials}
\eea
The pullback to $\Delta$ of the symplectic current will then be exact \cite{afk2000}, so that, using
Stokes theorem, the integral over $\Delta$ decomposes as in (\ref{deltaflux}).  The final symplectic structure
turns out to be
\bea
\bm{\Omega}_{\rm Cov}(\delta_1,\delta_2)
&=& \frac{1}{16 \pi G}\int_{M}
    \left[\delta_{1}A^{IJ} \wedge \delta_{2}\Sigma_{IJ}
    - \delta_{2}A^{IJ} \wedge \delta_{1}\Sigma_{IJ}\right]\nonumber\\
& &    + \frac{1}{4\pi} \int_M
    \left[\delta_{1}\bm{A} \wedge \delta_{2}(\star \bm{F})
    - \delta_{2}\bm{A} \wedge \delta_{1}(\star \bm{F})\right]\nonumber\\
& &
    -\frac{1}{8\pi G}\oint_{\IHslice}
    \left[\delta_{1}\psi \delta_{2}\areaform
    - \delta_{2}\psi \delta_{1}\areaform \right]
    - \frac{1}{4\pi} \oint_{\IHslice}
    \left[
     \delta_{1}\chi \delta_{2}(\star \bm{F})
    - \delta_{2}\chi \delta_{1}(\star \bm{F}) \right] \; .\nonumber\\
\label{fullsymplectic}
\eea

\subsubsection*{Symmetry classes and the phase space of rigidly rotating horizons}

For the purposes of black hole mechanics and the quantum theory of black
holes, it is useful to categorize different black hole geometries according to their
symmetry groups.  A \textit{symmetry} of an isolated horizon is an infinitesimal diffeomorphism
on $\IH$ which preserves $q_{ab}$ and $D$, and at most rescales $\ell$ by a positive
constant. (We restrict consideration to infinitesimal symmetries because the symmetry vector
fields we consider will not always be complete.)  By definition of an isolated horizon,
diffeomorphisms generated by $\ell$ are symmetries, so that the symmetry group $G_\IH$  of $\IH$
is at least
1-dimensional.  But beyond this, the group of symmetries of an isolated horizon, unlike
the symmetries of the asymptotic-flat metric at spatial infinity, is not universal.  The
symmetry group can be classified according to its dimension into three categories:
\begin{enumerate}
\item Type I: $(q,D)$ is spherically symmetric, and $G_\IH$ is four-dimensional.
\item Type II: $(q,D)$ is axisymmetric, and $G_\IH$ is two-dimensional.
\item Type III: $(q,D)$ has no symmetry other than $\ell$, and $G_\IH$ is
one-dimensional.
\end{enumerate}
Note that these symmetries refer \textit{only} to geometric structures intrinsic to the horizon.
No assumption is made about symmetries of fields outside the horizon, even in an arbitrarily small
neighborhood of $\Delta$.

Type II horizons are the most interesting ones, because they allow rotation
and distortion, but still have enough structure to admit a clear notion of (quasi-local)
angular momentum, as we shall see.  They include the Kerr-Newman family of horizons as well as generalizations
possessing distortion due to external matter, other black holes, or even `hair' due to matter
such as Yang-Mills for which the black hole uniqueness theorem does not apply. In the rest of
this section, we will focus on the type II case.

Let us fix an axial vector field
$\phi$ on $\Delta$ such that it commutes with any one and hence all
members of $[\ell]$. Define ${\bm \Gamma}^\phi$ to be the set of
all data $(e, A, {\bm A})$ in ${\bm \Gamma}_{\rm IH}$ such that
(1.) the induced
$q_{ab}$ and $\D$ on $\Delta$ possess $\phi$ as a symmetry:
$\pounds_\phi q_{ab} = 0$, and $[\pounds_\phi, D_a] = 0$, and
(2.) $\pounds_\phi \star \pb{\bm{F}} = \pounds_\phi \pb{\bm{F}} = 0$.
%
%
Endow ${\bm \Gamma}^\phi$ with the pull-back, via inclusion, of the symplectic structure on
${\bm \Gamma}_{\rm IH}$ derived above --- we denote the resulting symplectic structure by ${\bm \Omega}^\phi$.
%
%
The resulting phase space $({\bm \Gamma}^\phi, {\bm \Omega}^{\phi})$ is called
the \textit{phase space of rigidly rotating horizons}.  This will form the basis of the discussion for the
rest of this section.  Furthermore, whenever a partial Cauchy surface $M$ is used in the following, we assume
it is chosen such that its intersection $\IHslice$ with $\Delta$ is everywhere tangent to $\phi$.
%
%

\subsection{Conserved charges and the first law}

\subsubsection*{Symmetries and Hamiltonian flow.}

In $\bm{\Gamma}^\phi$, all space-times possess as isolated horizon symmetries the fixed vector fields
$\phi$ and $\ell$. A general symmetry at $\Delta$ thus takes the
form
\bea
W^a \hateq \B{W} \ell^a + \A{W} \phi^a \, ,
\label{gensymm}
\eea
where $\B{W}$ depends on which $\ell\in[\ell]$ is used.
Both $\B{W}$ and $\A{W}$ are constants on $\Delta$, but may be
`q-numbers' --- i.e., may vary on \textit{phase space}.
Suppose we extend $W^a$ arbitrarily to the rest of the space-time,
such that at spatial infinity it approaches a symmetry of the asymptotic flat metric
there.  One can then ask: Is the
flow generated by $W^a$ Hamiltonian?
Because the flow generated by $W^a$ preserves the boundary conditions at
$\Delta$ and at infinity involved
in the definition of ${\bm \Gamma}^\phi$, there exists a corresponding well-defined
variation $\delta_{W}$ on ${\bm \Gamma}^\phi$.
The flow will be Hamiltonian if and only if there exists a Hamiltonian $H$
on ${\bm \Gamma}^\phi$ such that $\delta H = {\bm \Omega}^{\phi}(\delta, \delta_W)$ for all variations $\delta$.
Using (\ref{fullsymplectic}), explicitly this becomes
\bea
\delta H = {\bm \Omega}^{\phi}(\delta, \delta_W) &=& \frac{-1}{8\pi G} \oint_{\IHslice} \A{W}\delta[(\phi \lrcorner \omega) \areaform]
             + \kappa_{(W)} \delta(\areaform)\nonumber\\
         & & - \frac{1}{4\pi} \oint_{\IHslice} \A{W} \delta[(\phi \lrcorner {\bm A})
             \star {\bm F}] - \Phi_{(W)} \delta(\star {\bm F})\nonumber\\
         & & + \frac{1}{16\pi G} \oint_{\IHslice_{\infty}} \tr[\delta A \wedge (W \lrcorner \Sigma)
             + (W \lrcorner A) \delta \Sigma]\nonumber\\
         & & - \frac{1}{4\pi} \oint_{\IHslice_\infty} \delta {\bm A} \wedge (W \lrcorner \star {\bm F})
             + (W \lrcorner {\bm A}) \delta(\star {\bm F}) \; ,
\label{omega_genw}
\eea
where
$\kappa_{(W)} = W^a \omega_a$ and $\Phi_{(W)} := - W^a \bm{A}_a$
respectively denote the surface gravity and electric potential
`in the frame defined by $W$'.
%
%
%
%

In the following, we will find such an $H$ first for
$W$ a \textit{pure rotation}, and then
an appropriate \textit{time translation}.
This will lead to the definition of the horizon angular momentum,
horizon mass, and a proof of a version of the first law of black hole
mechanics involving \textit{only quasi-locally defined quantities}.
In passing, the definition of angular momentum multipoles and the associated mass multipoles
will also be introduced.

\subsubsection*{Horizon angular momentum as generator of rotations}

Angular momentum is the generator of rotations.
In terms of equation (\ref{gensymm}),
the case of spatial rotations corresponds to $\B{W} = 0$ and $\A{W} = 1$,
in which case equation (\ref{omega_genw}) reduces to
\begin{displaymath}
\bm{\Omega}^{\phi}(\delta, \delta_\phi)
= -\frac{1}{8\pi G} \oint_{\IHslice} \delta[(\phi \lrcorner \omega)
\areaform]
- \frac{1}{4\pi} \oint_{\IHslice} \delta[(\phi \lrcorner \bm{A})
\star \bm{F}] - \delta J_{\rm ADM} \, ,
\end{displaymath}
so that the boundary integral at infinity is exact, equal to
the variation of the ADM angular momentum.
One can see that the terms associated with $\IHslice$ also
form an exact variation.  If we set
\begin{equation}
J_\Delta := - \frac{1}{8\pi G} \oint_{\IHslice} (\phi \lrcorner \omega)
\areaform - \frac{1}{4\pi} \oint_{\IHslice} (\phi \lrcorner \bm{A})
\star \bm{F} \, ,
\label{Jdef}
\end{equation}
then
\begin{equation}
\bm{\Omega}^\phi(\delta, \delta_W)
= \delta(J_\Delta - J_{\rm ADM}) \, ;
\end{equation}
this means that $J_\Delta - J_{\rm ADM}$ is the Hamiltonian generating rotations in
$\bm{\Gamma}^\phi$. Whereas $J_{\rm ADM}$ represents the angular momentum of the entire
space-time, $J_\Delta$ can be interpreted as the angular momentum
of the black hole itself, so that $J_{\rm ADM} - J_\Delta$ is then the angular momentum outside
of the horizon.
Note how the extension of $\phi^a$ in the bulk between $\Delta$
and infinity did not matter precisely because the expression for the symplectic
structure consists only in boundary terms.

One can show that the gravitational contribution to the horizon angular momentum (that is, the first term in (\ref{Jdef}))
is equivalent to the Komar integral.
To show that this statement is true, note that the normalization $\ell^{a}n_{a}=-1$
implies that the rotation one-form is given by
$\omega_{a}=-n^{b}\nabla_{a}\ell_{b}=\ell^{b}\nabla_{b}n_{a}$.  Substituting this into
the first term in (\ref{Jdef}), integrating by parts and using
the Killing property of $\phi$, we have:
\bea
J_{\rm Grav} &=& \frac{1}{8\pi G}\oint_{\IHslice}(\phi \lrcorner \nabla_{\ell}n)\areaform
                     = -\frac{1}{8\pi G}\oint_{\IHslice}(\nabla_{\ell}\phi \lrcorner n)\areaform\nonumber\\
                     &=& -\frac{1}{16\pi G}\oint_{\IHslice}(\ell \lrcorner d\phi) \lrcorner n\areaform
                     = -\frac{1}{8\pi G}\oint_{\IHslice}\star\d\phi \; .
\eea
This is the Komar integral for the gravitational contribution to the angular momentum of $\Delta$, evaluated at $\IHslice$.
Note that $J_{\rm Grav}$ is equivalent to the Komar integral even in the presence of Maxwell fields.

\subsubsection*{Multipoles}

The gravitational angular momentum can also be written as a particular moment of the \textit{imaginary part of $\Psi_2$}.
This then leads to a way to
define higher order angular momentum \textit{multipoles} \cite{aepv2004}.
%
%
Let us see
how this comes about.  $\phi$ is a symmetry of the
intrinsic geometry of $\Delta$ and therefore also a symmetry of $\areaform$.
This implies
$\pounds_{\phi}\areaform \hateq\d(\phi\lrcorner\areaform)\hateq0$ so that
$\phi\lrcorner\areaform= \d g$ for some smooth function $g$.  Fix the freedom to add a constant to $g$ by
imposing $\oint_{\IHslice} g \areaform = 0$, so that $g$ is unique.  Some manipulation then
gives
\bea
J_{\rm Grav} &=& -\frac{1}{8\pi G}\oint_{\IHslice}(\phi \lrcorner \omega)\areaform
                       = -\frac{1}{8\pi G}\oint_{\IHslice}(\d\omega) g
                       = -\frac{1}{4\pi G}\oint_{\IHslice} g {\rm Im} \Psi_2 \areaform\; ,
\label{closedforms}
\eea
where (\ref{curvatureofomega}) has been used in the last line. One can show that $g$ always has the range
$(-R_\Delta^2, R_\Delta^2)$ where $4 \pi R_\Delta^2 := \area_\Delta$ is the areal radius of the horizon.  In fact, there exists a canonical round metric $q^o_{ab}$ determined by the axisymmetric metric $q_{ab}$, sharing the same area element and Killing field $\phi$.  In terms of
the standard spherical coordinates $(\theta, \phi)$ adapted to $q^o_{ab}$
(with $\phi^a = \left(\frac{\partial}{\partial \phi}\right)^a$),
one has $g=R_\Delta^2 \cos \theta$, and the angular momentum can be written
\begin{equation}
J_{\rm Grav} = -\frac{R_\Delta^4}{4\pi G} \oint_{\IHslice} (\cos \theta) \Psi_2 \d \Omega
\; ,
\end{equation}
where $\d \Omega = \sin \theta \d \theta \wedge \d \phi$.  This leads to a more general definition of angular momentum \textit{multipoles}
\begin{equation}
J_n := -\frac{R_\Delta^{n+3}}{4\pi G} \oint_{\IHslice} P_n(\cos \theta){\rm Im}\Psi_2 \d \Omega \; ,
\end{equation}
where $P_n$ are the Legendre polynomials, and where
the power of $R_\Delta$ is chosen to give the correct dimensions.
One can similarly take moments of the scalar curvature $\mathcal{R}$ of the 2-metric on any slice of the horizon to give
the \textit{mass} multipoles
\begin{equation}
M_n := \frac{M_\Delta R_\Delta^{n+2}}{8\pi G} \int_{\IHslice} P_n(\cos \theta) \mathcal{R} \d \Omega \; ,
\end{equation}
where $M_\Delta$ is the horizon mass, to be derived in the next section.
These two sets of multipoles satisfy $J_0 = 0$ (no angular momentum monopole), $J_1 = J_\Delta$,
$M_0 = M_\Delta$, and $M_1 = 0$ (one is `automatically in the center of mass frame').
More importantly, the multipoles are \textit{diffeomorphism invariant}
and together \textit{uniquely determine both $q_{ab}$ and the horizon derivative operator $D_a$ up to diffeomorphism} \cite{aepv2004}.
They have been used both in numerical relativity \cite{skb2006, owen2009, saijo2011}, as well as to extend the black hole entropy calculation
in loop quantum gravity (to be reviewed later in this chapter) to include rotation and distortion compatible with
axisymmetry \cite{aev2004}.

\subsubsection*{Horizon energy and the first law}

Energy is the generator of time-translations.  To derive a notion of energy, we therefore seek
a linear combination of the symmetry vector fields $\ell$ and $\phi$
\bea
t \hateq B_{(t,\ell)}\ell - \Omega_{(t)}\phi \,
\label{tathorizon}
\eea
to play the roll of time translation,
such that the corresponding flow is Hamiltonian.
Here, as in the case of the Kerr-Newman family of solutions,
$\Omega_{(t)}$ is interpreted as the \textit{angular velocity of the horizon} relative to $t$.
The generator of this flow will
then provide us with the horizon energy.
In turns out that, in order to accomplish this, unlike for the case of rotational symmetry
considered above,
the coefficients $B_{(t, \ell)}$, $\Omega_{(t)}$, and hence the vector field
$t^a$, \textit{must be allowed to vary from point to point in phase space.}
One can see a hint that this would be necessary already in the Kerr-Newman family of solutions:
the stationary Killing vector field $t$, determined by the condition that it approach
a fixed unit time-translation at infinity,
as a linear combination of $\ell$ and $\phi$,
is not constant over the family. For example, in the Reissner-Nordstr\"om sub-family of solutions,
$\Omega_{(t)}$ is zero, and otherwise it is not.

Fix a unit time
translation field of the asymptotic flat metric.
At each point of $\bm{\Gamma}^\phi$, we introduce a vector field $t$
on the entire space-time such that (1.)
$t \= B_{(t,\ell)}\ell-\Omega_{(t)}\phi$ at $\Delta$ and
(2.) $t$ approaches the fixed unit time translation at infinity.
Evaluating the symplectic structure (\ref{omega_genw}) at $(\delta,\delta_{t})$, one obtains
\bea
\label{omegat}
\bm{\Omega}^\phi (\delta,\delta_{t}) = \bm{\Omega}_{\Delta}(\delta,\delta_{t}) + \delta E_{\rm ADM} \, ,
\eea
where $E_{\rm ADM}$ is the usual ADM energy, given by an integral at spatial infinity $\IHslice_\infty = M \cap \mathcal{I}$.
The integral at
$\Delta$ is given by
\begin{eqnarray}
\bm{\Omega}_{\Delta}(\delta, \delta_t)
= \frac{\kappa_{(t)}}{8\pi G}\delta\!\oint_{\IHslice}\!\!\!\areaform
  + \frac{\Phi_{(t)}}{8\pi G}\delta\!\oint_{\IHslice}\!\!\!\star\bm{F}
  + \frac{\Omega_{(t)}}{8\pi G}\delta\!\oint_{\IHslice}\!\!\!\left[(\phi \lrcorner \omega)\areaform
  + (\phi\lrcorner\mathbf{A})\star\bm{F}\right],
  \label{finalomega}
\end{eqnarray}
where we used $\kappa_{(t)}=\pounds_{t}\psi=t\lrcorner\omega$ and $\Phi_{(t)}=\pounds_{t}\chi=t\lrcorner\bm{A}$.
The flow determined by $t$ will be Hamiltonian iff $\bm{\Omega}^\phi (\delta, \delta_t)$  is an exact variation.
From (\ref{omegat}) and (\ref{finalomega}), this will be the case iff
there exists a phase space function $E_{\Delta}$ such that for all variations $\delta$,
\bea
\delta E_{\Delta}^{(t)}
\!\!\!&=&\!\!\! - \frac{\kappa_{(t)}}{8\pi G}\delta\!\oint_{\IHslice}\!\!\!\areaform
    - \frac{\Phi_{(t)}}{8\pi G}\delta\!\oint_{\IHslice}\!\!\!\star\bm{F}
    - \frac{\Omega_{(t)}}{8\pi G}\delta\!\oint_{\IHslice}\!\!\!\left[(\phi \lrcorner \omega)\areaform
    + (\phi\lrcorner\mathbf{A})\star\bm{F}\right] .
\label{edelta}
\eea
If this is true, one has
\bea
\label{tgen}
\bm{\Omega}^\phi (\delta, \delta_t) = \delta(E_{\rm ADM} - E_{\Delta}) \; .
\eea
The above equation tells us that, if $E_{\Delta}$ exists, $E_{\rm ADM} - E_{\Delta}$ will be the hamiltonian
generating the flow determined by $t$. $E_{\rm ADM}$ is interpreted as the energy of the
entire space-time, whereas $E_{\Delta}$ will have the interpretation of the energy of the black hole
proper.  $E_{\rm ADM} - E_{\Delta}$ is therefore the energy of the gravitational radiation and matter
present in the entire intervening region between the horizon and infinity.
%
%

We now ask: What are the conditions which $t$ must satisfy in order for $E_{\Delta}$ to exist?  How many such vector
 fields $t$ are there? This will be clarified in the following
proposition. We shall see that there are an infinite number of evolution vectors
leading to a Hamiltonian flow, and hence for which $E_{\Delta}$ exists.

%
\noindent{\bf Proposition 4.}
\emph{There exists a function $E_{\Delta}$ such that (\ref{edelta}) holds if and only if $\kappa_{(t)}$, $\Phi_{(t)}$
and $\Omega_{(t)}$ can be expressed as functions of the `charges' $\area_\Delta$,
$Q_\Delta$ $J_\Delta$ defined by}
\bea
\area_{\Delta}     &=& \oint_{\IHslice}\areaform
\label{entropy6}\\
Q_{\Delta}     &=& \frac{1}{8\pi G}\oint_{\IHslice}\star\bm{F}
\label{charge}\\
J_{\Delta}     &=& \frac{1}{8\pi G}\oint_{\IHslice}\left[(\phi \lrcorner \omega)\areaform
    + (\phi\lrcorner\mathbf{A})\star\bm{F}\right] \, ,
\label{angularmomentum}
\eea
\emph{and satisfy the integrability conditions}
\bea
\frac{1}{8\pi G}\pdiff{\kappa_{(t)}}{J_{\Delta}} = \pdiff{\Omega_{(t)}}{\area_{\Delta}} \, ,
\qquad
\frac{1}{8\pi G}\pdiff{\kappa_{(t)}}{Q_{\Delta}} = \pdiff{\Phi_{(t)}}{\area_{\Delta}} \, ,
\qquad
\pdiff{\Omega_{(t)}}{Q_{\Delta}} = \pdiff{\Phi_{(t)}}{J_{\Delta}} \; .
\label{integrability}
\eea
\noindent{\bf Proof.}
Let $\dbldif$ and $\dblwedge$ denote exterior derivative and exterior product on
the infinite dimensional space
$\bm{\Gamma}^{\phi}$.  Suppose there exists a phase space function
$E_{\Delta}$ such that
(\ref{edelta}) holds.
Equation (\ref{edelta}) is equivalent to
\bea
\dbldif E_{\Delta} = \frac{\kappa_{(t)}}{8\pi G} \dbldif \area_{\Delta}
                     + \Phi_{(t)}\dbldif Q_{\Delta} + \Omega_{(t)} \dbldif J_{\Delta} \; .
\label{edelta2}
\eea
Because the gradient of $E_{\Delta}$ is a linear combination of the gradients
of $\area_{\Delta}$, $Q_{\Delta}$, $J_\Delta$, $E_\Delta$
is a function only of these parameters,
$E_{\Delta} = E_{\Delta}(\area_{\Delta}, Q_{\Delta}, J_\Delta)$.
%
%
%
%
The chain rule then implies
\bea
\dbldif E_{\Delta} = \pdiff{E_{\Delta}}{\area_{\Delta}} \dbldif \area_{\Delta}
                     + \pdiff{E_{\Delta}}{Q_{\Delta}} \dbldif Q_{\Delta}
                     + \pdiff{E_{\Delta}}{J_{\Delta}} \dbldif J_{\Delta} \; .
\label{chainrule}
\eea
From, for example, the Kerr-Newman family of solutions, we know $\area_{\Delta}$, $Q_{\Delta}$ and $J_{\Delta}$ are
independent quantities and hence $\dbldif \area_{\Delta},
\dbldif Q_{\Delta},\dbldif J_{\Delta}$ are linearly independent.
From (\ref{edelta2}) and (\ref{chainrule}) one then has
\bea
\label{thermopartials}
\frac{\kappa_{(t)}}{8 \pi G} = \pdiff{E_{\Delta}}{\area_{\Delta}}, \qquad
\Phi_{(t)} = \pdiff{E_{\Delta}}{Q_{\Delta}}, \qquad
\Omega_{(t)} = \pdiff{E_{\Delta}}{J_{\Delta}} \; ,
%
%
\eea
which imply in turn that $\kappa_{(t)}$, $\Phi_{(t)}$, and $\Omega_{(t)}$ similarly depend only on
$\area_{\Delta},Q_{\Delta}$,and $J_{\Delta}$.
Commutativity of partials together with (\ref{thermopartials}) then imply the integrability conditions
(\ref{integrability}).

Conversely, if $\kappa_{(t)}, \Phi_{(t)}, \Omega_{(t)}$ depend on $\area_\Delta$, $Q_\Delta$, and $J_\Delta$
alone and satisfy the integrability conditions (\ref{integrability}), there exists $E_{\rm ADM}$ such that (\ref{thermopartials}),
and hence such that (\ref{edelta2}) and (\ref{edelta}) hold.  \qedd

If we replace the surface gravity $\kappa_{(t)}$ and the area $\area_\Delta$
with the Hawking temperature $T_{(t)} = \kappa_{(t)}/2\pi$ and
the Bekenstein-Hawking entropy $S_{\Delta} = \area_\Delta/4G$, (\ref{edelta})
becomes
\begin{equation}
\delta E_{\Delta} = T_{(t)} \delta S_{\Delta} + \Phi_{(t)}\delta Q_{\Delta}
                          + \Omega\delta J_{\Delta} \; .
\end{equation}
which is none other than the first law of black hole mechanics.  One thus sees that
\textit{the necessary and sufficient condition for $t$ to be Hamiltonian is that there
exist a phase space function $E_\Delta$ such that the first law holds.}

One can show that there are an infinite number of possible functions $\kappa_{(t)}$, $\Phi_{(t)}$, $\Omega_{(t)}$
of $a_\Delta$, $Q_\Delta$, and $J_\Delta$ satifying the integrability conditions (\ref{integrability}).  For each of these, there is
a corresponding $t^a$ which, by the foregoing proposition, is hamiltonian, and hence gives rise to a notion of horizon
energy $E_\Delta$.  That there are an infinite number of hamiltonian time evolution
vector fields $t^a$ and corresponding notions of horizon energy is not so surprising: In the context of stationary space-times,
the only thing which allows one to isolate a single time-translation vector field $t$
at the horizon is the global stationarity of the space-time, which rigidly connects
the choice of $t$ at the horizon to the choice of $t$ at infinity, where it can be fixed by
requiring it to be a time translation of the asymptotic Minkowski metric.
In the present context of isolated horizons,
by contrast, one does not have global stationarity.  Thus, physically, one expects an ambiguity in $t$
and hence in the definition of horizon energy.  What is remarkable is that \textit{all} of these horizon energies satisfy the first law.

Furthermore,
the foregoing proposition gives one tight control over
this infinity of possible evolutions and corresponding energies.
In the following subsection, we will see how this control can be exploited to select a \textit{unique, canonical}
$t$ on the entire phase space, and hence a \textit{unique, canonical} notion of horizon energy, which one
calls the horizon mass.

\subsubsection*{Mass}

For practical applications, such as in numerical relativity, it is useful to isolate
a canonical notion of horizon mass.  In the context of Einstein-Maxwell theory,
the uniqueness theorem provides a way to do this.  Specifically, one takes advantage of the
fact that, although the phase space $\bm{\Gamma}^\phi$ is far larger than the
Kerr-Newman family of solutions, the Kerr-Newman family nevertheless forms a subset of
$\bm{\Gamma}^\phi$, and on this subset we can stipulate that $t$ be equal to the standard
stationary Killing field
$t$ for the Kerr-Newman space-time in question, determined in the usual way by requiring $t$
to approach a unit time translation at infinity. Combined with the results of Proposition 4,
\textit{this suffices to uniquely determine $t$ on the entire phase
space $\bm{\Gamma}^\phi$}.

Let us see how this comes about.  From Proposition 4, $\kappa_{(t)}$ must be a function
of $\area_{\Delta},Q_{\Delta},J_{\Delta}$ alone.
The fact that we have stipulated that $t$ and hence $\kappa_{(t)}$ take their
standard canonical values on the Kerr-Newman family uniquely fixes this function to be
\begin{equation}
\kappa_{(t)} =
\kappa_{0}(\area_{\Delta},Q_{\Delta}, J_{\Delta})
:= \frac{R_{\Delta}^4 - G^2(Q_{\Delta}^2 + 4J_{\Delta}^2)}
    {2R_{\Delta}^3\sqrt{(R_{\Delta}^2 + G Q_{\Delta}^2)^2 + 4G^2 J_{\Delta}^2}} \, ,
\end{equation}
where $R_{\Delta}=\sqrt{\area_{\Delta}/(4\pi)}$ is areal radius of $\IHslice$.
With $\kappa_{(t)}$ uniquely determined, the integrability conditions (\ref{integrability})
can be used to uniquely determine the associated $\Phi_{(t)}$ and $\Omega_{(t)}$ as well.
One finds that these are given by
\bea
\Phi_{(t)} &=& \frac{Q_{\Delta}(R_{\Delta}^{2} + G Q_{\Delta}^{2})}
             {R_{\Delta}\sqrt{(R_{\Delta}^2 + G Q_{\Delta}^2)^2 + 4G^2 J_{\Delta}^2}} \, , \\
\Omega_{(t)} &=& \frac{2G J_{\Delta}}{R_{\Delta}\sqrt{(R_{\Delta}^2 + G Q_{\Delta}^2)^2
         + 4G^2 J_{\Delta}^2}} \; .
\eea
Finally, equation (\ref{thermopartials}) can be integrated to uniquely determine the horizon energy up to an additive constant.
This additive constant can be fixed by requiring the horizon energy to be equal to the usual value of the energy
when evaluated on members of the Kerr-Newman family.  The resulting horizon energy we denote $M_\Delta$ and is called
the \textit{horizon mass}.  It is given by
\begin{equation}
M_{\Delta} = \frac{\sqrt{(R_{\Delta}^2 + GQ_{\Delta}^2)^2 + 4G^2 J_{\Delta}^2}}{2GR_{\Delta}} \; .
\label{horizonmass}
\end{equation}
Note that this definition of the horizon mass involves only quantities \textit{intrinsic} to the horizon.
This is in contrast to the ADM or Komar definitions of mass, in which reference to infinity is required.

Finally, note that this strategy for selecting a canonical notion of mass works in Einstein-Maxwell theory
only because the uniqueness theorem holds.
If we consider theories in which the uniqueness theorem
fails to hold, such as Einstein-Yang-Mills theory, it is no longer possible to consistently stipulate that
$t^a$ (and hence $\kappa_{(t)}$, $\Phi_{(t)}$,and $\Omega_{(t)}$) be equal to its canonical value when evaluated on stationary solutions: Because each triple $(\area_\Delta, J_\Delta, Q_\Delta)$ corresponds to multiple stationary solutions, such a prescription now would contradict the requirement of Proposition 4 that $\kappa_{(t)}$, $\Phi_{(t)}$, and $\Omega_{(t)}$ must be functions of $\area_\Delta, J_\Delta, Q_\Delta$ alone. In such theories, one still has an infinite family of first laws.  It is just that one cannot select a single one as canonical.

\section{Quantum isolated horizons and black hole entropy}

%
%

As first observed by Bekenstein \cite{bekenstein1973}, if we identify the surface gravity $\kappa$ of a black hole with its temperature
  and its area $\area$ with its entropy, then
the first law of black hole mechanics takes the form of the standard first law of thermodynamics.
With this identification, the zeroeth and second laws of black hole mechanics, reviewed in the introduction,
furthermore take the form of the standard zeroeth and second laws of thermodynamics and furthermore provide
a way to preserve the second law of thermodynamics in the presence of a black hole.
These observations led Bekenstein to hypothesize that a black hole
is a thermodynamic object, its area \textit{is} its entropy, and its surface gravity \textit{is} its temperature.
This hypothesis was impressively confirmed by Hawkings calculation showing that black holes
radiate thermally at a temperature equal to $T = \hbar \kappa/2\pi$, a calculation which also fixed the
coefficient relating $T$ and $\kappa$, which in turn allowed one, via the first law, to fix the coefficient relating
the entropy $S$ to the area $\area$:
\begin{displaymath}
S = \frac{\area}{4 \ell_{\rm Pl}^2} \; .
\end{displaymath}
This is the celebrated Bekenstein-Hawking entropy.  It provides tantalizing evidence that black holes are thermodynamic
objects with an entropy whose statistical mechanical origin lay in quantum theory.  To account for the statistical mechanical
origin of black hole entropy has become one of the challenges for any quantum theory of gravity.

In this section we will
review such an account provided by \textit{quantum isolated horizons} using
the approach to quantum gravity known as \textit{loop quantum gravity}.  The key principal which guides
loop quantum gravity is that of \textit{background independence}, which is equivalent to the symmetry principle
of \textit{diffeomorphism covariance} at the foundation of classical general relativity.

For simplicity, we here review the more recent $SU(2)$ covariant account of this derivation of the entropy \cite{enp2009, enpp2010},
which built to a great extent on the original work \cite{abck1997, ack1999, abk2000} which used a $U(1)$ partial gauge-fixing to aide calculations.
Furthermore, we will only present the \textit{type I} case --- i.e., the case in which the intrinsic geometry of the
horizon is spherically symmetric.  The type II and type III horizons are handled in
\cite{aev2004, be2010}; see also \cite{pp2010}.  Lastly, again in order to focus on
the key ideas, we will handle only pure gravity.  Inclusion of Maxwell and Yang-Mills fields, and even non-minimally coupled scalar fields
can also be handled, and can be found in the references \cite{abk2000, acs2003, ac2003}.

\subsection{Phase space and symplectic structure}

In loop quantum gravity, one uses the \textit{Ashtekar-Barbero} variables \cite{barbero1995}, which
we use in the form
$(\gA, \Sigma)$ consisting in the Ashtekar-Barbero connection
$\gA^i_a$, and
the `flux-2-form' $\Sigma^i_{ab}$, where $i = 1,2,3$.
These are related to the space-time variables used in the prior sections via
\begin{eqnarray*}
\gA^i &=& -\frac{1}{2} \epsilon^i{}_{jk} \undersquigarrow{A}^{jk} + \beta \undersquigarrow{A}^{i0}\\
\Sigma^i &=& \epsilon^i{}_{jk} \undersquigarrow{e}^j \wedge \undersquigarrow{e}^k \, ,
\end{eqnarray*}
where the arrows $\undersquigarrow{}$ denote pull-back to the partial Cauchy surface $M$,
$\epsilon^i{}_{jk}$ denotes the three dimensional alternating tensor,
and $\beta \in \R^+$ is the \textit{Barbero-Immirzi parameter},
a quantization ambiguity consisting in a single real number.
Furthermore, in the above equation, it is understood that the space-time
variables $A_a^{IJ}, e_a^I$ are in the \textit{time-gauge} in which
$e^a_0$ is normal to the partial Cauchy surface $M$, reducing the internal
local gauge group from $SO(1,3)$ to $SO(3)$.
%
%
%
%
To give a sense of the meaning of these variables, in
terms of generalized ADM variables, one has
$\Sigma^i_{ab} = \undertilde{\epsilon}_{abc} (\det e^d_j)^{-1} e^{ci}$
where $e^a_i$ is the triad, and
the two terms in the expression for $\gA$ are given by
$ -\frac{1}{2} \epsilon^i{}_{jk} \undersquigarrow{A_a}^{jk} = \Gamma^i_a$, the spin connection
determined by the triad, and $\undersquigarrow{A}^{i0} = K^i_a = K_{ab} e^{bi}$
where $K_{ab}$ the extrinsic curvature of $M$.

Let $\bm{\Gamma}^{(I)}$ denote the space of possible fields $(\gA, \Sigma)$ on $M$
satisfying Einstein's equations, such that the inner boundary $\IHslice:= \Delta \cap M$ corresponds to a type I isolated horizon of a
\textit{fixed} area $\area_0$, and such that appropriate asymptotically flat boundary conditions
are satisfied at infinity.  Thus, the intrinsic geometry of $\IHslice$ is restricted to be round in $\bm{\Gamma}^{(I)}$.
%
%
%
If one starts from the symplectic structure (\ref{fullsymplectic}), imposes the above conditions, and rewrites the
result using the Ashtekar-Barbero variables $(\gA^i, \Sigma^i)$, one obtains the following symplectic structure:
%
%
\bea
\label{entropysymplectic}
\bm{\Omega}^{(I)}(\delta_1, \delta_2)
&=& \frac{1}{8\pi G \beta}\int_{M}[\delta_1 \Sigma^i \wedge \delta_2 \gA_i - \delta_2 \Sigma^i \wedge \delta_1 \gA_i]\nonumber\\
& & - \frac{1}{8\pi G \beta}\frac{\area_o}{\pi(1-\beta^2)} \int_\IHslice \delta_1 \gA_i \wedge \delta_2 \gA^i \; .
\eea
In deriving this, one makes key use of the fact that, in terms of the Ashtekar-Barbero
variables, the isolated horizon boundary conditions take the form
\begin{equation}
\label{horbc}
\Sigma^i = -\frac{\area_o}{\pi(1-\beta^2)} F^i(\gA) \;
\end{equation}
where $F^i(\gA) = d \gA^i + \frac{1}{2}\epsilon^i{}_{jk} \gA^j \wedge \gA^k$ is the curvature of $\gA^i$.
This is referred to as the \textit{quantum horizon boundary condition}.
%
%
%
%
Also note that the symplectic structure (\ref{entropysymplectic}) consists in \textit{two terms} --- a bulk term equal
to the standard symplectic structure used in loop quantum gravity,
and a surface term at $\IHslice$,
equal to the symplectic structure of an $SU(2)$ \textit{Chern-Simons theory}.  This observation, together
with the horizon boundary condition, are the keys to the quantization of this system.

\subsection{Quantization}

%
%

As just remarked, the symplectic structure (\ref{entropysymplectic}) consists in two terms,
a bulk term identical to that used in LQG,
and a horizon, $SU(2)$ Chern-Simons theory term.  This motivates one to \textit{quantize the bulk and horizon
degrees of freedom separately}.  After they have been separately quantized, the horizon boundary condition
(\ref{horbc}) will be used to couple them again.  Let us start with the bulk.  The bulk Hilbert space of states
$\Hil^B$ is the standard one in loop quantum gravity, spanned by \textit{spin-network states}.
A spin-network state
$|\gamma, \{j_e\}, \{i_v\} \rangle$ is labelled by a collection of edges $\gamma$
called a graph with each edge labelled by
a half-integer spin $j$, and each point at the end of an edge, called a vertex, labelled by
an \textit{intertwiner} among the irreducible representations on the edges meeting at the vertex.
Spin-network states are eigenstates of \textit{area}: For any 2-surface $T$, one classically has a corresponding
area $\area_T$, and hence in quantum theory a
corresponding area operator $\hat{\area}_T$, and one has
\begin{equation}
\hat{\area}_T |\gamma, \{j_e\}, \{i_v\} \rangle = \left(8 \pi G \beta
\sum_{p \in T \cap \gamma} \sqrt{j_p(j_p +1)}\right) |\gamma, \{j_e\}, \{i_v\} \rangle \; .
\end{equation}
In the present case, because the spatial manifold $M$ has boundary $\IHslice$, $\gamma$ may have edges
which \textit{intersect} this boundary.  We call such intersections punctures.
Suppose we are given a finite set of points $\mathcal{P}$, and assignment of a spin $j_p$ to each point.
Let $\Hil^B_{\mathcal{P}, \{j_p\}}$ denote the span of all spin-network states whose punctures
are at the points
$\mathcal{P}$ and the spins on the corresponding edges are $\{j_p\}$.
%
%
In terms of these spaces, the full bulk Hilbert space can be expressed
as a direct sum over spins
and a direct limit over all finite subsets of $\IHslice$:
%
%
\begin{equation}
\Hil^B = \dirlim \bigoplus_{\{j_p\}_{p \in \mathcal{P}}} \Hil^B_{\mathcal{P}, \{j_p\}} \; .
\end{equation}

Consider one of these spaces $\Hil^B_{\mathcal{P}, \{j_p\}}$.
On this space, the action of the operator corresponding
to the two form $\Sigma^i_{ab}$ pulled back to $\IHslice$ reduces to \cite{al2004}
%
%
%
\begin{equation}
\epsilon^{ab} \hat{\Sigma}^i_{ab}(x)
= 8 \pi G \beta \sum_{p \in \mathcal{P}} \delta(x, x_p)
\hat{J}^i(p) \, ,
\end{equation}
where at each puncture $p$ the operators $J^i(p)$ satisfy the usual angular momentum algebra
$[J^i(p), J^j(p)] = \epsilon^{ij}{}_k J^k(p)$.
%
%
%
Substituting this into (\ref{horbc}), we get
\begin{equation}
\label{qhbc}
-\frac{\area_o}{\pi (1-\beta^2)} \epsilon^{ab} \hat{F}^i_{ab}
= 8 \pi G \beta \sum_{p \in \mathcal{P}} \delta(x, x_p)
\hat{J}^i(p) \; .
\end{equation}
Here $\hat{F}^i_{ab}$ is the quantization of the curvature of
$\gA^i_a$ pulled back to $\IHslice$, and hence the curvature of
the $SU(2)$ \textit{Chern-Simons connection}.
This shows us that the generators $\hat{J}^i(p)$
act as point sources for the $SU(2)$ Chern-Simons theory.
The quantization of $SU(2)$ Chern-Simons theory with such point
sources is well-understood --- see for example Witten \cite{witten1989}.
In the end, for each fixed set of spins $j_p$ associated to the point
sources, one obtains a Hilbert space of states $\Hil_k^{CS}(\{j_p\})$,
where $k:= \area_o/(2\pi \beta (1-\beta^2) \ell_P^2)$ is the `level' of the Chern-Simons theory,
appearing in
the coefficient of the surface symplectic structure in (\ref{entropysymplectic}).
This Hilbert space can be viewed as a subset of the tensor product of carrying spaces associated to the $SU(2)$ representations labeled by
the spins $j_p$
\begin{displaymath}
\Hil_k^{CS}(\{j_p\}) \subset \otimes_{p \in \mathcal{P}} V_{j_p} \, ,
\end{displaymath}
where $\hat{J}^i(p)$ acts on each $V_{j_p}$ irreducibly.
$\Hil^B_{\mathcal{P}, \{j_p\}}$ can similarly be decomposed in a way that makes the
action of the generators apparent
\begin{displaymath}
\Hil^B_{\mathcal{P}, \{j_p\}} = \Hil_{\{j_p\}} \bigotimes \left(\otimes_{p \in \mathcal{P}} V_{j_p} \right) \, ,
\end{displaymath}
where again $V_{j_p}$ denotes the spin $j_p$ carrying space on which $\hat{J}^i(p)$ acts irreducibly.
For a given set of punctures $\mathcal{P}$ and spins $\{j_p\}$, one is therefore led to the following space of states satisfying the quantum
version of the horizon boundary condition (\ref{horbc}),
\begin{equation}
\Hil^{\Kin}_{\mathcal{P}, \{j_p\}} = \Hil_{\{j_p\}} \otimes \Hil_{k}^{CS}(\{j_p\}) .
\end{equation}
Upon reassembling these spaces using a direct sum over spins and direct limit over sets of punctures,
one obtains the full space of states solving the horizon boundary condition (\ref{qhbc}):
\begin{equation}
\Hil^{\Kin} = \dirlim \bigoplus_{\{j_p\}_{p \in \mathcal{P}}} \Hil_{\{j_p\}} \otimes \Hil_{k}^{CS}(\{j_p\}) \; .
\end{equation}
Here `Kin' indicates that the diffeomorphism and Hamiltonian constraints have not yet been imposed.
%
%
Imposition of the diffeomorphism constraint roughly speaking leads to replacement of
quantum states with their
\textit{diffeomorphism equivalence class}.
See \cite{abk2000} for further details.
The solution to the diffeomorphism constraint
then takes the form
\begin{equation}
\Hil^\Diff = \bigoplus_n \bigoplus_{(j_p)_1^n} \Hil_{(j_p)_1^n} \otimes \Hil_{k}^{CS}(\{j_p\}) \, ,
\end{equation}
where now one is no longer summing over possible positions of punctures, but only over the total \textit{number} of
punctures and corresponding spins.  Because lapse is restricted to vanish on the horizon \cite{ack1999, enpp2010},
the Hamiltonian constraint is imposed only in bulk, resulting in replacement of $\Hil_{(j_p)_1^n}$
with an appropriate subspace $\tilde{\Hil}_{(j_p)_1^n}$, yielding the final space of physical states
\begin{equation}
\Hil^\Phys = \bigoplus_n \bigoplus_{(j_p)_1^n} \tilde{\Hil}_{(j_p)_1^n} \otimes \Hil_{k}^{CS}(\{j_p\}) \; .
\end{equation}
For further details, we refer to \cite{abk2000, enpp2010}.
%
%
%

\subsection{Ensemble and entropy}

%
%
%
%

We are interested in the ensemble of physical states in $\Hil^\phys$ consisting in
eigenstates of the horizon area $\hat{\area}_\IHslice$ with eigenvalue in the range
$(\area_o - \delta, \area_o + \delta)$ for some tolerance $\delta$.  Let $\Hil^\bh$
denote the span of such states.  One would like to define the entropy via the standard
Boltzmann formula as the logarithm of the dimension of this space.  However, the dimension
of this space is infinite, for the simple reason that all of bulk degrees of freedom are
represented in it --- degrees of freedom which are irrelevant for the question
which interests us.

To eliminate this infinity, one needs to talk more precisely about
the horizon degrees of freedom.  To this end, we define
%
%
\begin{equation}
 \Hil^S := \bigoplus_{n} \bigoplus_{(j_p)_1^n} \Hil_{k}^{CS}(\{j_p\}) \; .
\end{equation}
We say that a given horizon state $\psi^S \in \Hil_{k}^{CS}(\{j_p\}) \subset \Hil^S$
is `compatible' with $\Hil^\bh$ if there exists $\psi^B \in \tilde{\Hil}_{(j_{p})_1^n}$ such
that $\psi^B \otimes \psi^S \in \Hil^\bh$. Let $\Hil_S^\bh$ denote the span of all such
compatible horizon states.  The entropy is then given by the usual Boltzmann
%
%
formula
%
%
\begin{displaymath}
S := \log \dim \Hil_S^\bh \; .
\end{displaymath}
One finds \cite{abbdv2009, enpp2011} the final entropy to be
\begin{equation}
S = \frac{\beta_0 \area_o}{4\beta \ell_P^2}, \qquad \beta_0 = 0.274067\dots
\end{equation}
so that if we choose $\beta = \beta_0 = 0.274067\dots$, the Bekenstein-Hawking entropy formula results.
Note the role played by $\beta$: it is a single quantization ambiguity, which in principle
can be fixed by a single experiment.  By requiring the Bekenstein-Hawking entropy law to hold
for any black hole with spherical intrinsic geometry (type I)
and a single value of the area, $\beta$ becomes fixed.  The fact that the entropy law then continues to be satisfied for
black holes of other areas is already a non-trivial test.
But, in fact, the present framework has been shown to pass much stronger tests as well:
if one extends the frame-work to include arbitrary Maxwell and Yang-Mills fields
\cite{abk2000}, or type II (axisymmetric) and type III (generic) isolated horizons \cite{aev2004, be2010},
one again obtains an entropy equal to one quarter times the area,
for the \textit{same} value of the Barbero-Immirzi parameter.\footnote{As long
as one identifies the horizon degrees of freedom in the same way as has been done here, counting the
$j$ labels as distinguishing horizon states. See \cite{gm2008,gm2011} for a discussion.}
One can even include a scalar field which is non-minimally coupled to gravity \cite{acs2003, ac2003}.
As mentioned in the introduction, this leads to a modified entropy law.
The present framework, when extended to include a non-minimally coupled scalar field, has been shown to
lead precisely to the required modified entropy law, again for the same value of $\beta$.

\section{Summary and discussion}

We have reviewed in detail the mathematical foundations of IHs in classical and quantum gravity.  Let us
briefly summarize the main features of this framework.

An IH is defined to be a null hypersurface with compact spatial cross-sections, whose outgoing null normal is non-expanding, and on which
certain components of the Levi-Civita connection are Lie dragged by the null normal.
We have seen that a well-defined first-order action principle can be given for space-times possessing an IH as an inner boundary,
with the IH boundary conditions ensuring that the action is functionally differentiable.
From second variations of this action, a conserved symplectic
current can be identified on the \emph{covariant phase space} of
solutions to the field equations.  Using this, we have seen how one can construct a well-defined canonical framework.
In this framework, local rotations and time translations at the horizon are generated by clear notions of
angular momentum and energy of the IH.  These quantities are independent of the ADM charges at infinity,
and satisfy a quasi-local version of the first law of black hole mechanics. Furthermore, the expression for the horizon angular momentum
can be naturally generalized to give a quasi-local, diffeomorphism-invariant notion of angular momentum  and mass
multipoles.

When the symplectic structure of this canonical framework is recast in terms of real $SU(2)$
Ashtekar connections and triads, the boundary term in the symplectic structure at the horizon
becomes that of \textit{$SU(2)$ Chern-Simons theory}.  This leads one to
quantize the bulk degrees of freedom using \textit{loop quantum gravity}
methods, and the horizon degrees of freedom using
a well-understood quantization of $SU(2)$ Chern-Simons theory.
The bulk and surface degrees of freedom are then coupled through a quantum version of a
`horizon boundary condition' embodying the fact that the inner boundary is an isolated horizon.
The statistical entropy of an ensemble of states with area in a small window of values
is found to be equal to $1/4$ times the horizon area for a single, universal value of the Barbero-Immirzi
parameter, a single quantization ambiguity present in the loop quantization of gravity.

There are some key features of the IH framework that we would like to highlight at this point.
First, we stress that, in the IH framework, there is \emph{no need to make reference to asymptotic infinity
at all}.  This paradigm shift is essential in order that we may understand situations in which a black
hole is in equilibrium with dynamical fields in the exterior region that are in an arbitrarily small
neighbourhood of the horizon.
Moreover, we note that the isolated horizon definition places a restriction only geometric structures \textit{intrinsic to the horizon}.
Not only does this make the definition simpler to check in practice,
but it ensures that the \textit{full} complement of degrees of freedom outside the horizon remains intact.
This is important in quantum theory, and allows the methods of loop quantum gravity to be used outside the horizon without change.

Lastly, perhaps the most notable feature of the IH framework is that it provides a unified mathematical
construction for understanding equilibrium black holes in both classical and quantum gravity.  Indeed, the
IH framework can be used to study the geometry, topology, supersymmetry, mechanics and quantum statistical
mechanics of quasi-local black holes in general relativity (and beyond).

\section*{Acknowledgements}

We are grateful to Christopher Beetle for a careful reading of the manuscript for this chapter.
This work was supported in part by the
Natural Sciences and Engineering Research Council of Canada (TL), United States NSF grant PHY-1237510 (JE),
and by NASA through the University of Central Florida's NASA-Florida Space Grant Consortium (JE).


\noindent{Department of Physics, Florida Atlantic University, Boca Raton, FL 33431, U.S.A.}\\
\emph{Electronic mail: jonathan.engle@fau.edu}

\vspace{3mm}

\noindent{Department of Mathematical and Statistical Sciences, University of Alberta, Edmonton,
AB T6G 2G1, Canada}\\
\emph{Electronic mail: tliko@math.ualberta.ca}

\end{document}